\documentclass[12pt]{article}

\usepackage{newtxtext,newtxmath}

\usepackage{graphicx}
\usepackage[letterpaper,margin=1in]{geometry}

\renewenvironment{abstract}
	{\quotation}
	{\endquotation}

\date{}

\makeatletter
\renewcommand{\fnum@figure}{\textbf{Figure \thefigure}}
\renewcommand{\fnum@table}{\textbf{Table \thetable}}
\makeatother

\usepackage{scicite}

\usepackage{url}

\def\scititle{{\fontsize{22pt}{20pt}\selectfont
Future Amplification of Moist Weather Extremes in the Midlatitudes
}}

\title{\bfseries \boldmath \scititle}

\author{
	Funing~Li$^{1\ast}$,
	Talia~Tamarin-Brodsky$^{1}$\and
	\small$^{1}$ Earth, Atmospheric, and Planetary Sciences, Massachusetts Institute of Technology, Cambridge, MA, USA.\and
	\small$^\ast$Corresponding author. Email: lifuning1991@gmail.com
}

\begin{document} 

\maketitle

\begin{abstract} \bfseries \boldmath 
SIGNIFICANCE: Compound extremes involving humid heatwaves and severe convection are growing climate risks in the midlatitudes, yet the physical mechanism for their projected intensification is unclear. Here we combine theory, climate model simulations, and targeted experiments to show that projected changes in low-level atmospheric inversions govern the coupled changes in extremes of moist heat convective instability. A major pathway is amplified warming over upstream highland carried downstream by prevailing westerlies, which strengthens inversions and raises attainable moist heat and storm potential. This explains the emerging hotspots found over eastern North America and northeastern Asia. By linking thermodynamic constraints with large-scale dynamics, our results provide a unified framework for the intensity and spatial distribution of midlatitude moist extremes under climate change.
\end{abstract}  

\newpage
\begin{abstract} \bfseries \boldmath 
ABSTRACT: Moist heatwaves and convective storms frequently co-occur, posing compound risks. Although historically concentrated in the tropics, these moist weather extremes are projected to intensify substantially towards the midlatitudes, with regions downstream of major highland terrains, including northeastern Asia and eastern North America, emerging as hotspots of future change. Yet their physical drivers remain uncertain. Here we show that the intensification of concurrent moist heat and convection extremes in the midlatitudes is tightly constrained by changes in low-level atmospheric inversions. Specifically, we find that amplified warming over western highlands is transported downstream by prevailing westerlies, strengthening low-level thermal inversions and raising the attainable maxima of moist heat and convection. Targeted model experiments confirm the critical role of orographically elevated heating in driving these extremes. Our results reveal a mechanistic pathway for compound extremes and highlight low-level inversions as a key factor for emerging midlatitude risks of moist heat and severe weather under climate change.
\end{abstract}  

\section*{INTRODUCTION}

Heat and convection are fundamental atmospheric processes that give rise to a wide range of high-impact weather and climate hazards, including humid heatwaves and severe convective storms \cite{morss_etal_2011,ummenhofer_etal_2017}. Moist heat, arising from the combined effects of high temperature and humidity near the ground, exerts direct physiological stress on the human body \cite{sherwood_huber_2010,baldwin_etal_2023}, while moist convection, characterized by the rapid ascent of near-surface warm and humid air, produces heavy rainfall with widespread societal consequences \cite{stevens_2005atmospheric,bao_etal_2024}. These moist weather extremes constitute major and escalating climate threats under ongoing global warming \cite{coumou_Rahmstorf_2012}. 

Historically, moist heat and deep convection have been concentrated in low latitudes, where abundant surface moisture and heating supplies favor their occurrence \cite{matthews_2018,zipser_liu_2021}. Yet, these moist weather extremes in the midlatitudes exhibit substantially greater complexity owing to diverse impacts, from regional human activities \cite{kang_eltahir_2018,zhang_prein_2025}, heterogeneous land surface processes \cite{kong_huber_2023,li_etal_2024}, and large-scale atmospheric dynamics \cite{tamarin_etal_2020,jiang_etal_2025}. Although humid heat events are increasingly observed outside the tropics \cite{raymond_etal_2020,rogers_etal_2021}, assessments of heat extremes there have largely focused on dry-bulb temperature extremes \cite{domeisen_etal_2023,barriopedro_etal_2023}. Insufficient consideration of humidity during heatwaves has been shown to notably underestimate human heat stress in some midlatitude regions \cite{romps_lu_2022,lanzante_2024new,xu_etal_2025}. Meanwhile, midlatitude continents preferentially host some of the most intense convective storms on Earth \cite{Zipser_etal_2006} due to the episodic coexistence of high atmospheric instability and strong vertical wind shear \cite{Brooks_etal_2003}, which produce damaging winds \cite{prein_2023}, large hail \cite{tang_etal_2019}, and tornadoes \cite{li_etal_2024} that represent a dominant class of hazardous weather \cite{BillionDollar_2025}. Beyond their individual impacts, extreme moist heat and severe convective storms frequently co-occur across the midlatitudes \cite{sauter_etal_2023a,treppiedi_etal_2024}, such as southeastern Australia \cite{sauter_2023c,xu_etal_2023mega,ganguli_etal_2025}, Europe \cite{sauter_etal_2023b,vyshkvarkova_etal_2022}, China \cite{you_wang_2021,li_etal_2022heat,miao_etal_2024}, and the U.S. \cite{zhang_Villarini_2020,hao_etal_2024}. The co-occurrence of these extreme moist weather events synergistically amplifies human and socioeconomic damages over vast areas \cite{zhou_etal_2024heat,liu_etal_2025}.  

As the planet warms, moist heat \cite{russo_etal_2017}, convective weather \cite{Lepore_etal_2021}, and their compound occurrences \cite{jin_etal_2025} are projected to amplify substantially toward the midlatitudes, and might bring unprecedented humid heat stress and severe weather hazards to billions of people in regions with little prior exposure. In particular, several midlatitude regions, such as eastern portions of Eurasia and North America, are emerging as potential future hotspots for moist heat in a hotter climate \cite{coffel_etal_2017,vecellio_etal_2023,raymond_etal_2025}, where severe thunderstorm activity is also projected to expand markedly eastward \cite{Lepore_etal_2021}. Some of these projected changes have already been detected in observational records, including historical trends in tornadoes and large hail over the U.S. \cite{Gensini_Brooks_2018,tang_etal_2019} and in very large hail in China \cite{Battaglioli_etal_2025}. Despite these growing threats from moist weather extremes in the midlatitudes, the physical mechanisms governing their strength and future changes remain poorly understood. Unlike the tropics, where responses to warming are reasonably explained by fundamental thermodynamic principles \cite{zhang_etal_2021_heat,Singh_etal_2017}, they do not readily extend to the midlatitudes, where systematic departures from Clausius–Clapeyron scaling \cite{da_etal_2025,wang_moyer_2023} and quasi-equilibrium state \cite{zhang_2002convective,lafleur_etal_2023} have long been recognized. This mechanistic gap undermines confidence in projections of extreme weather \cite{sillmann_etal_2017} and heightens the risk of unforeseen climate impacts across densely populated midlatitude regions\cite{shaw_bjorn_2025}. 

As an effort to fill this gap, the recently developed inversion-constraint theory \cite{Li_Tamarin-Brodsky_2025}, which identifies low-level atmospheric inversions as a fundamental control on both moist heat and convection in the midlatitudes, offers a physically grounded framework for interpreting their future changes. The theory builds upon the prevailing hypothesis that the onset of deep convection terminates heatwaves \cite{buzan_huber_2020,zhang_boos_2023,duan_etal_2024}, but extends beyond the moist neutrality assumption commonly applied to quasi-equilibrium tropical convection \cite{emanuel_etal_1994,neelin_zeng_2000,singh_ogorman_2013} by explicitly incorporating the stored-energy process that characterizes continental severe convection in the midlatitudes \cite{Doswell_2001,lafleur_etal_2023,Tuckman_etal_2023,Tuckman_Emanuel_2024}. A low-level inversion features a shallow, stable layer above the surface in which (potential) temperature increases rapidly with altitude \cite{wood_Bretherton_2006}, with the strength defined by the maximum saturated moist static energy (MSE$^*_{max}$) at or above the lifting condensation level (LCL) \cite{Li_Tamarin-Brodsky_2025} (see $Materials$ \& $Methods$). Acting as an energy barrier that inhibits coupling between the near-surface and free-tropospheric atmosphere \cite{klein_etal_2018}, a strong low-level inversion permits abundant accumulation of near-surface heat and instability until the inversion is eroded and convection is triggered, and thus promotes the co-occurrence of moist heat and severe convection with their potential maxima set by the strength of the inversion \cite{Li_Tamarin-Brodsky_2025}. 

Here we investigate the future changes of concurrent moist heat and convection extremes across midlatitude continents to address two questions: what constrains amplifications of these moist weather extremes in a warmer climate, and what drives their emerging hotspots. Analyzing projections from the Coupled Model Intercomparison Project Phase 6 (CMIP6) \cite{Eyring_etal_2016}, we show that changes in midlatitude continental low-level inversions impose a tight physical constraint on the co-amplification of moist heat and convection under climate change. Specifically, the advection of elevated heating over high terrain, a primary pathway for generating downstream thermal inversions \cite{Carlson_etal_1983,Li_etal_2021}, intensifies under amplified upstream highland warming, thus strengthening downstream low-level inversions and preconditioning regions such as northeastern Asia and eastern North America for extreme moist heat and convection. We further conduct targeted climate model experiments to demonstrate the fundamental role of orographically elevated heating in driving midlatitude moist weather extremes. 

\section*{{RESULTS}} 

\subsection*{Future amplification of moist weather extremes constrained by inversions}

We first present projections of moist heat and convection extremes from a high-emission climate change scenario (ssp585), compared with historical simulations in CMIP6 models. Moist heat is quantified by near-surface moist static energy (MSE$_s$) \cite{Li_Tamarin-Brodsky_2025} that directly informs human heat stress, and the intensity of potential convection is measured by convective available potential energy (CAPE). These definitions are broadly consistent or interconvertible with commonly used metrics of humid heatwaves \cite{zhang_etal_2021_heat,raymond_etal_2021,kong_huber_2023,duan_etal_2024} and convective intensity \cite{Singh_etal_2017,Chen_etal_2020}. Given the co-occurring nature of extremes in moist heat and convection across midlatitude continents \cite{Li_Tamarin-Brodsky_2025}, we focus on compound extremes at each land grid point over the Northern Hemisphere midlatitudes, defined by the annual maximum MSE$_s$ and the associated CAPE (denoted as critical CAPE $\equiv$ CAPE$_c$ with critical implying at the time of annual maximum MSE$_s$). Here we present results based on four well-performing CMIP6 models, evaluated against the ERA5 reanalysis and selected for their relatively smaller biases, higher spatial correlations, and stronger physical consistency compared with the other five models (Supplementary Figs. \ref{fig_s01} and \ref{fig_s02}), but the rest of the models are shown in the Supporting Information (Supplementary Figs. \ref{fig_s03} and \ref{fig_s04}). See $Materials$ \& $Methods$ for details on definitions and assessments. 

\begin{figure} 
\centering
\centerline{\includegraphics[width=1.05\linewidth]{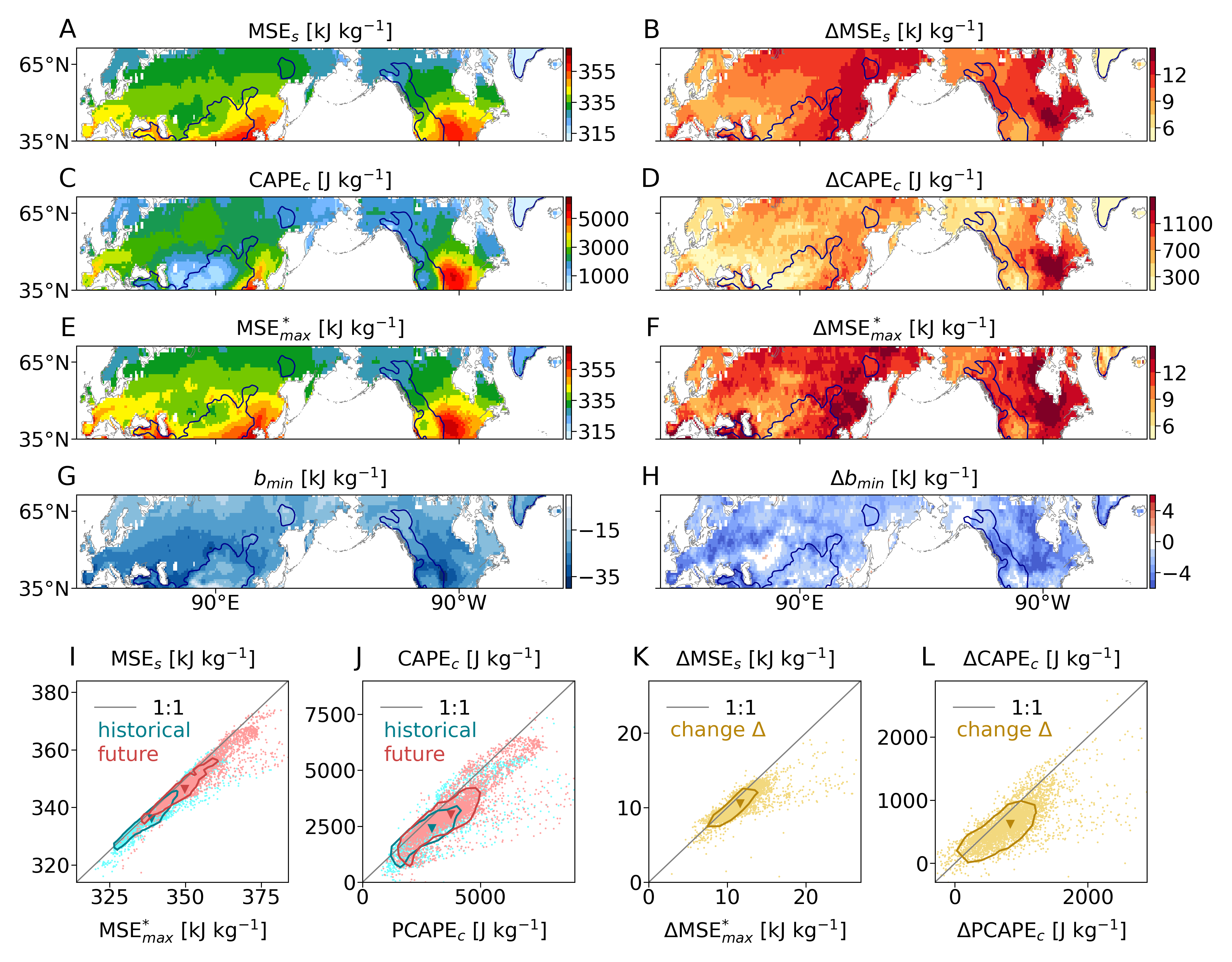}} 
\caption{\textbf{Inversion constraints on concurrent moist heat and convection projections.} 
(\textbf{A}): Historical (1980--2009) annual maximum near-surface moist static energy (MSE$_s$) and (\textbf{B}): its projected changes ($\Delta$MSE$_s$; 2065--2094 minus historical). (\textbf{C}--\textbf{H}): Same as (A--B) but for (C--D) potential concurrent convection (CAPE$_c$) defined as CAPE at the time of annual maximum MSE$_s$, (E--F) low-level inversion constraints (MSE$_{max}^{*}$) defined as the maximum saturated MSE in lower free troposphere at the time of annual maximum MSE$_s$, and (G--H) pre-existing inversion-induced energy barrier defined as the largest difference between MSE$_s$ and MSE$_{max}^*$ within five days prior annual maximum MSE$_s$. (\textbf{I}): Scatter plot of MSE$_{max}^*$ versus MSE$_s$ at its annual maxima for the historical (cyan colors) and future (red colors) periods, with each dot showing the multi-year and multi-model means at a midlatitude (35$^{\circ}$N--65$^{\circ}$N) land grid point, each closed line representing the 75\% probability density contour, and each triangle indicating the overall median value. (\textbf{J}): Same as (I) but for potential maximum CAPE (PCAPE$_c$) versus CAPE$_c$. (\textbf{K}--\textbf{L}): Same as (I--J) but showing projected changes for (K) $\Delta$MSE$_{max}^{*}$ versus $\Delta$MSE$_{s}$ and (L) $\Delta$PCAPE$_c$ versus $\Delta$CAPE$_c$. See $Materials$ \& $Methods$ for details of proxy definitions and model assessment.
}
\label{fig_01}
\end{figure}

Concurrent extremes of moist heat and potential convection exhibit an excess intensification over the midlatitude continents, further characterized by a pronounced west-east gradient with larger increases over the eastern halves of Eurasia and North America (Fig. \ref{fig_01}$B$ and $D$). Historically (Fig. \ref{fig_01}$A$ and $C$), the primary local maxima of midlatitude  MSE$_s$ and CAPE$_c$ are both centered over eastern China and central U.S., with a secondary local maximum over western Europe, consistent with ERA5 reanalysis (Supplementary Fig. \ref{fig_s01}$A$) and previous studies \cite{raymond_etal_2021,Taszarek_etal_2021_global,Li_Tamarin-Brodsky_2025}. While these regions continue to experience increasingly intense moist heat and convective environments in the future, the primary hotspots of projected amplification shift northeastward relative to the historical local maxima (Fig. \ref{fig_01}$B$ and $D$). Specifically, the dominant hotspots in Eurasia extend from northeastern China into eastern Russia, whereas in North America they are centered over the U.S. Midwest and expand northeastward to encompass a vast region of northeastern North America (Fig. \ref{fig_01}$B$ and $D$). Similar spatial patterns are also found in projections from lower-performing models (Supplementary Fig. \ref{fig_s03}$A$--$D$) and lower-resolution models (Supplementary Fig. \ref{fig_s04}$A$--$D$), although these models tend to exhibit larger amplification magnitudes that likely reflect their higher biases. These projected amplifications imply a substantial northeastward expansion of humid heatwave and severe storm risks in a warmer climate, with newly emerging hotspots exposing densely populated and socioeconomically important regions, consistent with prior evidence of increasing heat stress over North China Plain \cite{kang_eltahir_2018} and the U.S. Midwest \cite{vecellio_etal_2023}, as well as the potential eastward shift in U.S. severe storm activity \cite{Gensini_Brooks_2018,tang_etal_2019}. 

We further show that the future amplifications of midlatitude moist weather extremes are tightly constrained by projected changes in low-level atmospheric inversions, whose spatial intensification closely resembles the patterns of moist heat and convection changes (Fig. \ref{fig_01}$E$ and $F$). According to the inversion-constraint theory, the extent to which near-surface heat and convective instability can accumulate depends on the timing of convective initiation, which in turn is controlled by the erosion of the low-level inversion \cite{Li_Tamarin-Brodsky_2025}. These low-level inversions, characterized by the peak saturated moist static energy of the atmosphere within the lower free troposphere (MSE$^*_{max}$), constitute strong energy barriers (Fig. \ref{fig_01}$G$) that suppress convection and enable further buildup of near-surface moist heat and convective instability. Consequently, the strength of MSE$^*_{max}$ sets an upper bound on the attainable maximum intensity of moist heat and convection, represented by MSE$^*_{max}$ and PCAPE$_c$ on the right-hand sides of Eqs. \ref{eq1.1} and \ref{eq1.2} (see $Materials$ \& $Methods$). The widespread increase in MSE$^*_{max}$ in a warmer climate therefore indicates a strengthening of low-level inversions and a higher energetic threshold for convection initiation (Fig. \ref{fig_01}$H$), raising the upper limit on near-surface moist heat and potential convection.

Quantitatively, the inversion constraints (Eqs. \ref{eq1.1} and \ref{eq1.2}) are well reproduced by models, and more importantly, remain robust across climate states (Fig. \ref{fig_01}$I$ and $J$). Relative to the historical values, the future extremes shift systematically toward higher-value regimes but remain closely tied to the one-on-one line (Fig. \ref{fig_01}$I$ and $J$), indicating that projected increases in moist heat and convection extremes (i.e., $\Delta$MSE$_s$ and $\Delta$CAPE$_c$) are likewise constrained by the strengthening in low-level inversions (Fig. \ref{fig_01}$K$ and $L$). Therefore, at leading order, $\Delta$MSE$_s$ and $\Delta$CAPE$_c$ are determined by changes in inversion strength, with only small residual contributions:
\begin{equation}  
\label{eq2.1}
\Delta \mathrm{MSE}_s = \Delta \mathrm{MSE}^*_{max} + \Delta \epsilon_{MSE},
\end{equation}
\begin{equation}  
\label{eq2.2}
\Delta \mathrm{CAPE}_c = \Delta \mathrm{PCAPE}_c+ \Delta \epsilon_{CAPE}.
\end{equation}

Physically, the residual terms ($\epsilon_{MSE}$ and $\epsilon_{CAPE}$) are reflection of the remaining inversion barrier at the time of annual maximum moist heat, determining how closely the actual maxima approach the potential maxima set by the inversion constraint. The historical inversion residuals are negligible across most midlatitude continents (Supplementary Fig. \ref{fig_s05}$A$), consistent with the generally weak convective inhibition at the time of annual maximum moist heat (Supplementary Fig. \ref{fig_s05}$C$). Exceptions occur over southern Europe, the U.S. west coast and the Great Plains, where small negative residuals may imply the influence of external lifting forces that trigger convection earlier or imply insufficient near-surface heating and moistening \cite{Li_Tamarin-Brodsky_2025}. Future changes in the residuals ($\Delta \epsilon_{MSE}$ and $\Delta \epsilon_{CAPE}$) resemble their historical distributions and remain small and largely unchanged over most regions (Supplementary Fig. \ref{fig_s05}$B$). Slightly enhanced residual barriers, corresponding to enhanced convective inhibition during the peak moist heat (Supplementary Fig. \ref{fig_s05}$D$), are found over certain regions such as southern Europe and the U.S. Great Plains. Larger residuals may imply that severe convection could be harder to initiate and thus may lead to fewer but more intense storms in the future \cite{ashley_etal_2023,hill_etal_2025}. Alternatively, larger residuals could imply stronger external lifting forces that allow convection to initiate before the inversion is completely removed \cite{schumacher_Rasmussen_2020}. These regional responses warrant further investigations, but the projected residuals are otherwise very small over most of the Northern Hemisphere midlatitudes. 

Overall, we show that projected changes in low-level inversions serve as the dominant first-order control on the future intensification of concurrent moist heat and convection extremes in the midlatitudes. Consequently, further elucidating mechanisms governing inversion changes provides deeper physical insight into changes in compound moist weather extremes. In particular, we are interested in the prominent emerging hotspots over northeastern Asia and northeastern North America, and propose a dynamical explanation for the spatial footprint of extreme moist weather amplification across these regions.

\subsection*{Emerging future hotspots dynamically linked to amplified highland warming}

\begin{figure} 
\centering
\centerline{\includegraphics[width=1.05\linewidth]{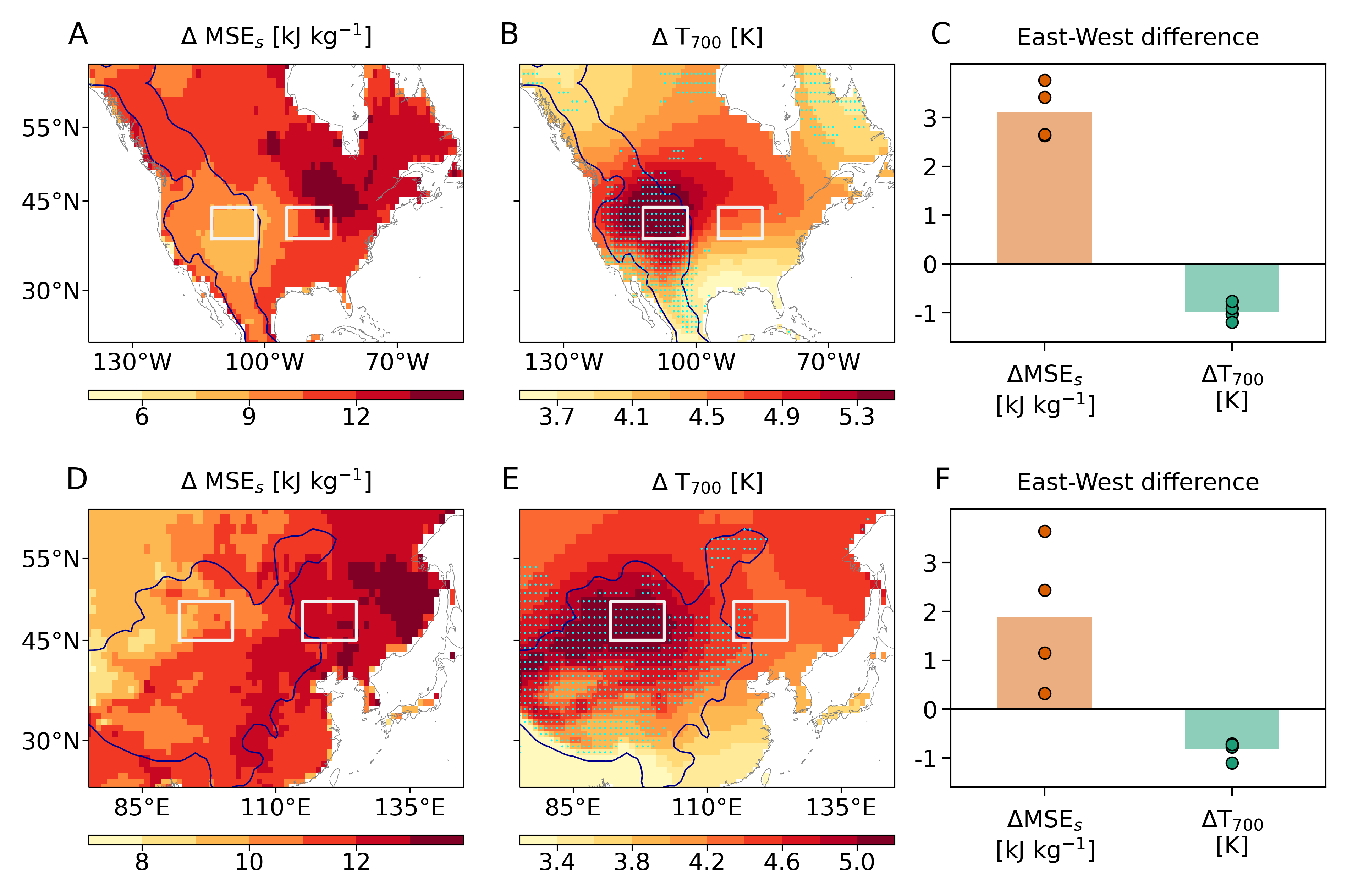}}
\caption{\textbf{Spatial contrasts in moist versus dry heat changes.} 
(\textbf{A}--\textbf{B}): Projected changes (future minus historical) of (A) annual maximum moist heat ($\Delta$MSE$_s$) and (B) summer mean 700-hPa air temperature ($\Delta T_{700}$) over North America. The dark blue line marks the 1000-m elevation contour. Two white boxes highlight sub-regions in the western highlands and the downstream U.S. Midwest. (\textbf{C}): Difference in $\Delta$MSE$_s$ and $\Delta T_{700}$ between the Midwest and western highland sub-regions, with bars showing multi-model means and filled circles indicating values from individual models. (\textbf{D}--\textbf{F}): Same as (A--C) but over eastern Asia, with the two sub-regions representing the western highlands and the downstream region over eastern Mongolia and northern China. Cyan dots in (B) and (C) indicate land grids primarily covered by bare sandy soil.
}
\label{fig_02}
\end{figure}

Geographically, emerging future hotspots of concurrent moist heat and convection extremes are concentrated over lowland regions downstream of major high terrain in North America and East Asia, with muted intensification over the highlands (Fig. \ref{fig_02}$A$ and $D$), producing a pronounced west-east gradient across both continents (Fig. \ref{fig_02}$C$ and $F$). In contrast, changes in dry-bulb air temperature, during the warm season when moist heat and convection often occur, exhibit the opposite zonal pattern with warming amplified over the western high terrain \cite{bishop_etal_2021,you_etal_2022}, which substantially enhances elevated heating in lower-to-middle atmosphere (e.g., 700 hPa; Fig. \ref{fig_02}$B$ and $E$). Over High Mountain Asia, this highland amplified warming may partly reflect elevation-dependent warming primarily associated with snow–albedo feedback \cite{palazzi_etal_2017,pepin_etal_2025}, although such mechanism is less likely to explain that over the Rockies, where elevation-dependent warming is mainly a cold-season phenomenon \cite{rangwala_etal_2016,minder_etal_2018}. Instead, because these high-terrain regions are dominated by drylands composed of arid sandy soils (dots in Fig. \ref{fig_02}$B$ and $E$) \cite{lian_etal_2021}, the projected soil-moisture drying over the drylands \cite{cook_etal_2020,lian_etal_2021}, which suppresses evaporative cooling and enhances sensible heating in a hotter climate \cite{mckinnon_etal_2024}, likely contributes to the amplified highland warming \cite{bauer_etal_2025}.

\begin{figure} 
\centering
\centerline{\includegraphics[width=1.05\linewidth]{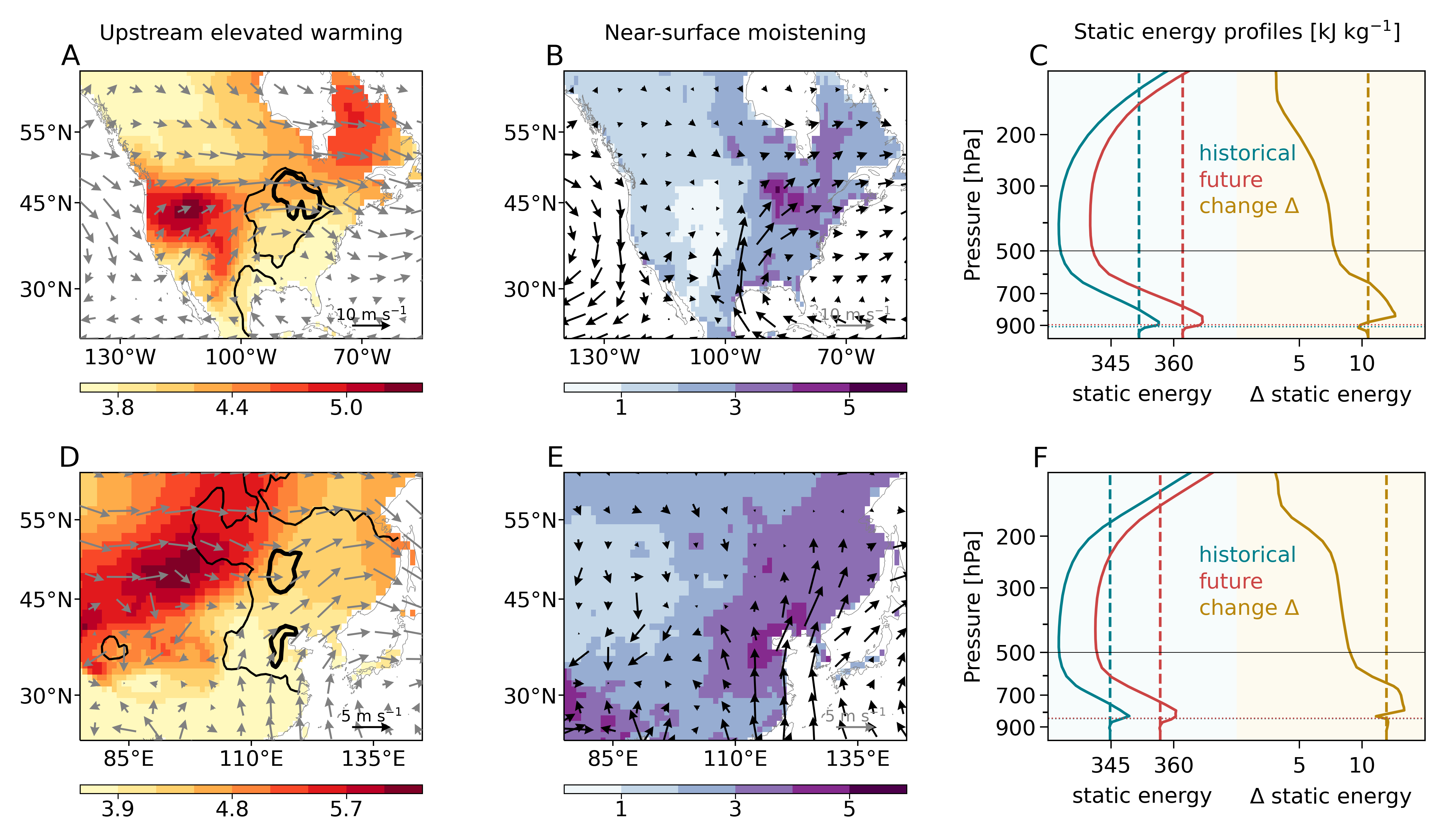}}
\caption{\textbf{Dynamical origins from upstream highland warming.}
(\textbf{A}): Projected changes (future minus historical) in composite $T_{700}$ (filled colors) and MSE$_{max}^{*}$ (black contours; thin at 12 kJ kg$^{-1}$, thick at 14 kJ kg$^{-1}$), and 700-hPa winds (vectors) from the future simulation, associated with extreme moist heat events over the U.S. Midwest. (\textbf{B}): Same as (A) but for projected changes in composite near-surface specific humidity ($q_s$; filled colors) and 925-hPa winds (vectors) from the future simulation. (\textbf{C}): Mean vertical profiles of environmental static energy (solid lines) and near-surface MSE$_s$ (dashed lines) for the historical, future, and change ($\Delta$; future minus historical) based on annual maximum MSE$_s$ cases over the U.S. Midwest land grids. The environmental static energy profile is defined as DSE($z$)+$L_vq_s$ for $z$ below lifting condensation level (LCL; thin dashed lines) and as MSE$^{*}(z)$ for $z$ above LCL. (\textbf{D}--\textbf{F}): Same as (A--C) but for cases over the northeastern Asia sub-region (e.g., eastern Mongolia and northern China). See $Materials$ \& $Methods$ for details of case selections and static energy profile definitions.
}
\label{fig_03}
\end{figure}

Further considering circulation dynamics in the midlatitudes, we hypothesize that amplified warming over upstream high terrain is a key driver of emerging future hotspots of moist weather extremes: under the prevailing midlatitude westerlies, enhanced warming over western highlands propagates eastward, intensifying low-level thermal inversions over the (north)eastern portions of the continents and thereby enabling continued accumulation of near-surface moist heat and convective instability in a warmer climate. To test the hypothesis, we perform composite analyses to explicitly demonstrate the dynamical pathway leading to extreme moist weather over the hotspot regions (see $Materials$ \& $Methods$). We focus on projections associated with extreme moist heat events over the U.S. Midwest (Fig. \ref{fig_03}$A$--$C$) and over eastern Mongolia-northern China (Fig. \ref{fig_03}$D$--$F$). For each continent, composite analyses for additional sub-regions farther east and northeast yield broadly consistent results (Supplementary Fig. \ref{fig_s09}). The composite extreme events are similar to the annual maxima of concurrent moist weather, as changes in composite mean MSE$_s$ and CAPE$_c$ peak over Midwest-northeast North America (Supplementary Fig. \ref{fig_s06}$A$ and $B$) and northeastern Asia (Supplementary Fig. \ref{fig_s06}$D$ and $E$), closely mirroring the projected intensification of annual maximum moist heat and convection across these regions (Fig. \ref{fig_01}$B$ and $D$).  

Our hypothesis is supported by composite patterns from projections, which show amplified warming of 700-hPa air temperature over the western highlands (filled color in Fig. \ref{fig_03}$A$ and $D$). Under strong midlatitude westerlies (vectors in Fig. \ref{fig_03}$A$ and $D$), these elevated warm anomalies are advected eastward to strengthen the formation of low-level inversions (MSE$^*_{max}$) across the U.S. Midwest and northeastern Asia (black contours in Fig. \ref{fig_03}$A$ and $D$). Notably, for composite events centered farther east and northeast, the highland warming is even stronger, thereby enabling the warm anomalies to propagate deeper into the downwind regions and to enhance inversions over much of northeastern North America and East Asia (Supplementary Fig. \ref{fig_s09}). 

The resulting intensification of low-level inversions, corresponding to larger pre-existing energy barriers to convection (Supplementary Fig. \ref{fig_s07}$A$ and $C$), raises the upper bound on near-surface moist heat and potential convection, which promotes the emergence of future moist weather hotspots by allowing excessive near-surface warming and moistening. While both near-surface air temperature and specific humidity are projected to increase, enhanced latent heat associated with rising humidity mainly contributes to the intensification of MSE$_s$ across the eastern halves of North America (Supplementary Fig. \ref{fig_s08}$A$ and $B$) and East Asia (Supplementary Fig. \ref{fig_s08}$C$ and $D$). The abundant surface humidity is associated with strong low-level southerly winds, which favor moisture transport from the warmer oceans into land regions (Fig. \ref{fig_03}$B$ and $E$), while local evapotranspiration from moist soils, vegetation canopies, and lakes may play an additional role \cite{Tuckman_etal_2023,zhang_etal_2025moisture}. Based on the scaling CAPE framework \cite{Li_Chavas_2021,wang_moyer_2023,Li_Tamarin-Brodsky_2025}, which approximates CAPE as proportional to the difference between MSE$_s$ and free tropospheric MSE$^*$ at 500 hPa (i.e., CAPE$=$0.22(MSE$_s-$MSE$^*_{500}$)), these near-surface energy accumulations also primarily contribute to the increased CAPE$_c$ (Supplementary Fig. \ref{fig_s06}$B$ and $E$), as increases in MSE$_s$ (Supplementary Fig. \ref{fig_s06}$A$ and $D$) over the hotspot regions substantially exceed increases in MSE$^*_{500}$ (Supplementary Fig. \ref{fig_s06}$C$ and $F$).

Projected changes in mean vertical profiles of static energy (Eq. \ref{eq5}) during annual maximum moist heat (see $Materials$ \& $Methods$), which directly diagnose buoyant instability of near-surface air parcels \cite{Li_Tamarin-Brodsky_2025}, further reveal the fingerprint of intensified low-level thermal inversions that strongly constrain the amplification of concurrent moist heat and convection (Eqs. \ref{eq2.1} and \ref{eq2.2}) over hotspots in North America (Fig. \ref{fig_03}$C$) and East Asia (Fig. \ref{fig_03}$F$). While the entire atmospheric column warms in the future climate (solid red versus blue lines in Fig. \ref{fig_03}$C$ and $F$), the largest increase occurs in the lower free troposphere between 700–900 hPa (solid orange lines in Fig. \ref{fig_03}$C$ and $F$), just above the lifting condensation level (zonal dashed lines in Fig. \ref{fig_03}$C$ and $F$), which indicates a pronounced strengthening of low-level inversions (MSE$^*_{max}$). The future increase in MSE$_s$ ($\Delta$MSE$_s$) is constrained by $\Delta$MSE$^*_{max}$ rather than by upper free tropospheric warming, as $\Delta$MSE$_s$ approaches $\Delta$MSE$^*_{max}$ but substantially exceeds middle-to-upper tropospheric (above $\sim$ 600-hPa height) $\Delta$MSE$^*$ (comparing dashed versus solid orange lines in Fig. \ref{fig_03}$C$ and $F$). Again, the larger increase in MSE$_s$ relative to MSE$^*_{500}$ accounts for the enhanced convective instability, while the residual inversion barrier that weakly increases in the future (i.e., slightly larger increases in MSE$^*_{max}$ than MSE$_s$ based on orange lines in Fig. \ref{fig_03}$C$ and $F$) reflects a weak increase in critical convective inhibition (in line with Supplementary Fig. \ref{fig_s05}$D$). The enhanced residual barrier is modest and confined to regions near dry high terrain (e.g., the U.S. Great Plains and interior East Asia) and Mediterranean regions, but remains negligible farther east of the continents (Supplementary Fig. \ref{fig_s07}$B$ and $D$), where greater soil moisture \cite{Emanuel_2023} and coastal moisture supply may make the inversion barriers easier to overcome. This result is also reflected in composite analyses over farther eastern and northeastern sub-domains within the hotspot regions (Supplementary Fig. \ref{fig_s09}), which show that future increases in MSE$_s$ are more tightly constrained by increases in MSE$^*_{max}$ (orange lines for static energy profiles in Supplementary Fig. \ref{fig_s09}), as compared to the interior regions (Fig. \ref{fig_03}$C$ and $F$).


\subsection*{Orographic elevated heating as a key driver of moist weather extremes}

At its core, the revealed dynamical pathway points to orographically elevated heating as a fundamental driver of downwind low-level inversions and associated extreme moist weather activity. To isolate this mechanism and substantiate our conclusion, we conduct numerical experiments that explicitly examine the response of midlatitude moist heat and convection extremes to elevated heating over upstream high terrain. The control simulation ($CTL$) is a historical global climate modeling for 1985--2014, and we obtain the net effect of elevated heating by comparing $CTL$ against an experiment in which highland surface heating is eliminated ($EXP$; See $Materials$ \& $Methods$ for experimental details).

\begin{figure}
\centering
\centerline{\includegraphics[width=1.05\linewidth]{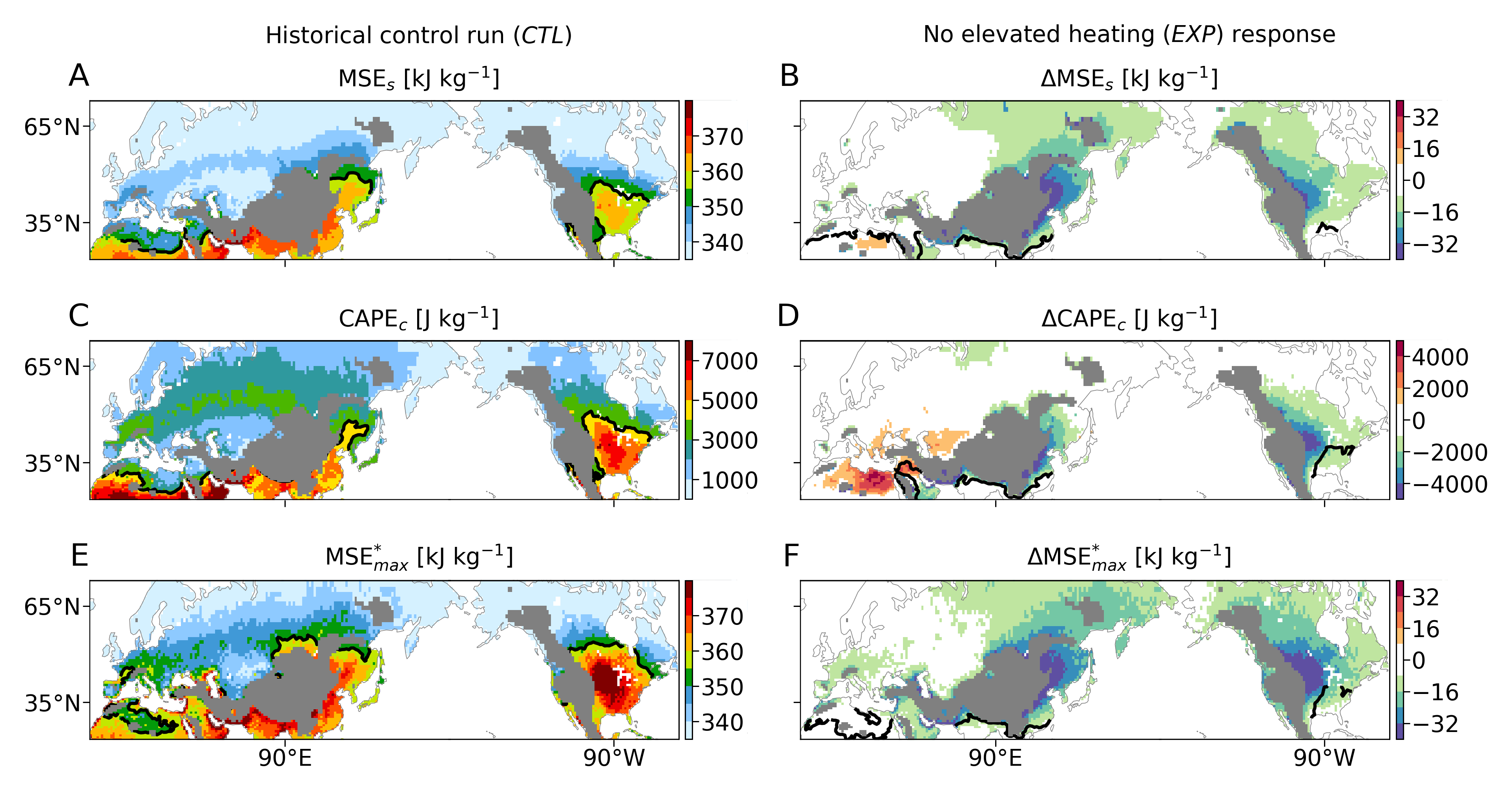}}
\caption{\textbf{Net responses to orographic elevated heating.} 
(\textbf{A}--\textbf{B}): Annual maximum MSE$_s$ from (A) the historical control simulation ($CTL$) and (B) changes in the experiment in which orographic elevated heating is removed ($EXP$ minus $CTL$). Black contour lines denote the extent of high MSE$_s$ values (MSE$_s=$355 kJ kg$^{-1}$) in (A) $CTL$ and (B) $EXP$, respectively. (\textbf{C}--\textbf{F}): Same as (A--B) but for (C--D) CAPE$_c$ with black contours representing CAPE$_c=$4000 J kg$^{-1}$, and (E--F) MSE$_{max}^{*}$ with black contours indicating MSE$_{max}^{*}=$355 kJ kg$^{-1}$. To emphasize responses over lowland regions, gray shading marks out highland surfaces above 800-m elevation, where the surface albedo is enhanced in $EXP$ (Supplementary Fig. \ref{fig_s10}$C$) to eliminate orographically elevated heating (Supplementary Fig. \ref{fig_s10}$F$). See $Materials$ \& $Methods$ for details of the experimental design.
} 
\label{fig_04}
\end{figure}

Compared with ERA5 reanalysis (Supplementary Fig. \ref{fig_s01}$A$) and the CMIP6 historical runs (Fig. \ref{fig_01}$A$ and $C$), the $CTL$ simulation reasonably reproduces the widespread distributions of annual maximum moist heat (MSE$_s$; Fig. \ref{fig_04}$A$) and concurrent potential convection (CAPE$_c$; Fig. \ref{fig_04}$C$) across subtropical and midlatitude land regions. Over the Afro-Eurasian continent, the primary local maximum extends from central Africa and the Indian Ocean coastal regions to northeastern Asia, while in North America the dominant local maximum is centered over central North America. Consistent with the inversion-constraint theory \cite{Li_Tamarin-Brodsky_2025}, these spatial patterns are broadly similar to the distribution of MSE$^*_{max}$ (Fig. \ref{fig_04}$E$), which acts as a strong lower-tropospheric energy barrier that tightly constrains the intensity of moist heat and convection. The inversion control is particularly robust in the midlatitudes due to the stored-energy nature of midlatitude continental convection \cite{Li_Tamarin-Brodsky_2025}, in line with the high convective inhibition established by inversions prior to the maximum moist heat (Supplementary Fig. \ref{fig_s10}$G$). 

Relative to the $CTL$, eliminating highland surface heating in the $EXP$ substantially suppresses extreme moist heat (Fig. \ref{fig_04}$B$) and severe potential convection (Fig. \ref{fig_04}$D$) across the midlatitudes. The most pronounced reductions occur downstream over eastern–northeastern Asia and the eastern half of North America, where the primary moist weather hotspots are nearly eliminated and high-value contours (black lines in Fig. \ref{fig_04}) of moist heat and convection both retreat significantly toward lower latitudes. These responses are associated with a remarkable weakening of low-level thermal inversions (Fig. \ref{fig_04}$F$), which reduces lower-tropospheric stability and inhibition (Supplementary Fig. \ref{fig_s10}$H$ and $I$), and thus lowering the upper limits on potential maximum moist heat and convection. These results directly reflect and reinforce the inversion-constraint mechanism and dynamical pathways of projected moist weather changes discussed in the previous sections. 

Although beyond the scope of this study, our experiments also indicate notable influences of orographically elevated heating on moist heat and convective environments in some subtropical regions, though changes across these regions are likely less directly relevant to inversion constraints \cite{Li_Tamarin-Brodsky_2025}. For example, eliminating large-scale elevated heating may contribute to the reduced MSE$_s$ and CAPE$_c$ along subtropical coastal areas (Fig. \ref{fig_04}$A$--$D$) by altering land-sea thermal contrast \cite{li_yanai_1996}, especially in monsoon-influenced areas such as South Asia and southwestern North America \cite{boos_kuang_2010,boos_pascale_2021}. Additionally, the $EXP$ shows enhanced CAPE$_c$ over North Africa (Fig. \ref{fig_04}$D$), a subtropical dryland region affected by anticyclonic circulation and midlevel warm advection by easterly trade winds from Asian highlands \cite{lu_etal_2018,chen_etal_2021TP}. Changes in this region are consistent with the CAPE scaling theory \cite{Li_Tamarin-Brodsky_2025} and observations \cite{Tuckman_Emanuel_2024}, suggesting that suppressing elevated heating may also increase convective instability in regions downstream of elevated terrain (under easterly trade winds) by cooling the free troposphere while leaving near-surface conditions relatively unchanged.  

\section*{{DISCUSSION}}

By demonstrating a unifying physical control on future changes in midlatitude moist heat and severe convection, our results place new emphasis on the role of low-level atmospheric thermal inversions in regulating extreme moist weather in a warming climate. While our analysis focuses on annual maximum moist weather, the broader intensification of inversions could have important implications for severe weather changes spanning the full distribution. Strong inversions serve as an attainable limit for extreme events but may remain a robust energy barrier for more typical conditions, such that enhanced inversions can intensify the most extreme convection while further suppressing weaker convection. This asymmetric influence of inversions favors fewer weak to moderate convection but more frequent intense convective storms under climate change, consistent with projected changes in the U.S. convective population \cite{rasmussen_etal_2020} and historical trends in severe convective events over Russia \cite{chernokulsky_etal_2022}. The resulting shift toward more intense and more frequent severe convective storms may further contribute to the disproportionate increase in extreme precipitation relative to light and mean precipitation over Europe \cite{da_etal_2025}, the United States \cite{cui_etal_2024}, and many other land regions \cite{dai_etal_2024}. 

The identified emerging hotspots indicate an expansion of extreme moist heat and severe convective storms toward regions where large-scale dynamics favor inversion intensification and low-level processes enable sufficient near-surface heating and moistening to approach and overcome the enhanced inversion limits in a warmer climate. However, in some regions, such as central North America and the Mediterranean region of Europe, inversions strengthen beyond the reach of even extreme moist heat events, thereby the initiation of even the most intense convection becomes increasingly difficult, which help explain the the observed \cite{Agee_etal_2016} and projected \cite{ashley_etal_2023} declines in supercell thunderstorm frequency over the U.S. Great Plains, as well as their projected reductions over the Mediterranean \cite{feldmann_etal_2025}. Better resolving these regional distinctions will require further investigation into changes in the governing processes of the surface energy budget \cite{rothlisberger_Papritz_2023} and changes in external lifting mechanisms \cite{weckwerth_Parsons_2006,bennett_etal_2006review}, which jointly determine how closely near-surface heat and instability can approach their theoretical limits. 

Together with many previous studies that have comprehensively investigated the roles of inversions in modulating surface radiation balance in the tropics \cite{clement_etal_2009,zhou_etal_2016} and polar regions \cite{medeiros_etal_2011,li_mace_2023}, as well as their impacts on air quality and pollution \cite{largeron_staquet_2016,sun_etal_2025inversion} and regional severe weather \cite{Ribeiro_Bosart_2018}, our efforts point toward the need for deeper mechanistic understanding and better model representation of low-level atmospheric stability to improve projections of future weather and climate. Our work establishes a dynamical linkage between changing patterns of extreme moist weather and elevated heating over upstream high terrain, reinforcing the geographical control of extreme weather on Earth \cite{baldwin_vecchi_2016,li_etal_2024,xie_etal_2025}. Changes in other processes such as large-scale subsidence \cite{neal_etal_2022} and radiative cooling \cite{Tuckman_Emanuel_2024} may also modulate low-level inversions and associated extremes, a topic that warrants further investigation. Furthermore, our results suggest that accurate simulation and prediction of moist heat and convection requires climate models to faithfully capture low-level inversions and the associated driving processes, with the inversion-constraint theory \cite{Li_Tamarin-Brodsky_2025} offering a physics-embedded pathway to evaluate model performance and constrain uncertainty in projections of weather and climate extremes.

\section*{MATERIALS AND METHODS}

\subsection*{Moist heat and potential convection}

We measure near-surface moist heat using near-surface moist static energy (MSE$_s$) \cite{Li_Tamarin-Brodsky_2025}, which is defined as the sum of near-surface sensible heat ($c_pT_s$), latent heat ($L_vq_s$), and geopotential energy ($gz_s$). Here, $c_{p}=1005$ J kg$^{-1}$ K$^{-1}$ is specific heat capacity of air at constant pressure, $L_{v}=2.5\times 10^6$ J kg$^{-1}$ is latent heat of vaporization, $g=9.81$ m s$^{-2}$ is gravitational acceleration, $T_s$ is 2-m air temperature, $q_s$ is 2-m specific humidity, and $z_s$ is height above sea level. Given its conservation under adiabatic processes, MSE$_s$ is thermodynamically identical to equivalent potential temperature \cite{madden_Robitaille_1970,betts_1974}, and is interconvertible with near-surface wet-bulb temperature \cite{Li_Tamarin-Brodsky_2025}. Similar to wet-bulb temperature, MSE$_s$ has been widely used as an indicator for human heat stress \cite{zhang_etal_2021_heat,raymond_etal_2021,duan_etal_2024}. In this study, we focus on extreme moist heat events in the midlatitudes, defined as the annual maximum MSE$_s$ over Northern Hemisphere continental regions.

Convective available potential energy (CAPE) is a key parameter for measuring the intensity of potential convection \cite{Singh_etal_2017}. Owing to the strong linear relationship between MSE$_s$ and CAPE, extreme moist heat events in the midlatitudes tend to co-occur with deep convection, with both typically maximizing at the onset of convection \cite{Li_Tamarin-Brodsky_2025}. This relationship is demonstrated by a scaling CAPE framework \cite{Li_Chavas_2021,wang_moyer_2023,Li_Tamarin-Brodsky_2025}, which provides a precise approximation of CAPE by the difference between MSE$_s$ and 500-hPa saturated MSE (MSE$_{500}^{*}$) scaled by a factor of $\alpha=$0.22 (see \cite{Li_Tamarin-Brodsky_2025} for detailed derivations and discussions). Accordingly, we also analyze CAPE at the time of annual maximum MSE$_s$ (i.e., critical CAPE, denoted as CAPE$_c$). CAPE$_c$ has been shown to largely coincide with the annual maximum CAPE over midlatitude continents \cite{Li_Tamarin-Brodsky_2025}. Consequently, future changes in CAPE$_c$ reflect changes in the intensity of severe convective weather and its co-occurrence with extreme moist heat events in the midlatitudes. We calculate CAPE for a near-surface air parcel by integrating the virtual temperature difference between the air parcel ($T_{v,a}$) and its environment ($T_v$) with respect to natural logarithm of pressure (ln$p$) from the level of free convection (LFC) to the equilibrium level (EL). Convective inhibition (CIN) is calculated analogously to CAPE, as the vertically integrated negative buoyancy from the surface to LFC.

\subsection*{Inversion and inversion-constraint theory}

The inversion constraint theory \cite{Li_Tamarin-Brodsky_2025} provides the theoretical foundation for this study. In the midlatitudes, the accumulation of near-surface moist heat is often accompanied by a pronounced buildup of convective instability, which can evolve beyond quasi-equilibrium due to the presence of pre-existing low-level energy inversions over continental regions \cite{Tuckman_Emanuel_2024,Li_Tamarin-Brodsky_2025}. The strength of these low-level inversions therefore constrains the maximum attainable intensities of both moist heat and potential convection \cite{Li_Tamarin-Brodsky_2025}. The key quantity of inversion, MSE$_{max}^*$, is defined as the maximum saturated moist static energy within the lower free troposphere, specifically at or above the lifting condensation level (LCL) but below 300 hPa. Acting as the maximum energetic barrier that near-surface air parcels must overcome to initiate deep convection, MSE$_{max}^*$ tightly limits the upper bounds of extreme moist heat and convective intensity \cite{Li_Tamarin-Brodsky_2025}, given by

\begin{equation}  
\label{eq1.1}
MSE_s\leq MSE^*_{max}
\end{equation}

\begin{equation}  
\label{eq1.2}
CAPE_c\leq PCAPE_c
\end{equation}
where $PCAPE_c=0.22(MSE^*_{max}-MSE^*_{500})$ is the potential maximum intensity of convection determined by inversion and 500-hPa saturated MSE based on the scaling CAPE relation \cite{Li_Tamarin-Brodsky_2025}. In this work, we apply the theory to future climate to determine the extent to which changes in low-level energy inversions ($\Delta$MSE$^*_{max}$) can predict changes in extreme moist heat ($\Delta$MSE$_s$) and potential convection ($\Delta$CAPE$_c$). 

\subsection*{CMIP6 climate simulations}

Low-level inversions and their constraints on heat and convection develop on synoptic timescales and exhibit pronounced diurnal variability, thereby high temporal resolution data are preferred for adequately capturing these processes. In addition, high vertical resolution is essential for resolving inversion structure and the buildup of instability. Accordingly, we analyze simulations from nine Coupled Model Intercomparison Project Phase 6 (CMIP6) models \cite{Eyring_etal_2016}, with each providing 6-hourly outputs on full model levels. These models (with Variant Label, horizontal grid size, and number of vertical levels in parentheses) are: CNRM-ESM2-1 (r1i1p1f2, 128$\times$256, 91) \cite{seferian_etal_2019}, CNRM-CM6-1 (r1i1p1f2, 128$\times$256, 91) \cite{voldoire_etal_2019}, MPI-ESM1-2-LR (r1i1p1f1, 96$\times$192, 47) \cite{mauritsen_etal_2019}, MPI-ESM1-2-HR (r1i1p1f1, 192$\times$384, 95) \cite{muller_etal_2018,mauritsen_etal_2019}, CMCC-CM2-SR5 (r1i1p1f1, 192$\times$288, 30) \cite{cherchi_etal_2019}, MRI-ESM2-0 (r1i1p1f1, 160$\times$320, 80) \cite{yukimoto_etal_2019}, MIROC6 (r1i1p1f1, 128$\times$256, 81) \cite{tatebe_etal_2019}, MIROC-ES2L (r1i1p1f2, 64$\times$128, 40) \cite{hajima_etal_2020}, and CanESM5 (r1i1p2f1, 64$\times$128, 49) \cite{swart_etal_2019}.

For each model, we compare the period of 1980--2009 from the historical run with the period of 2065--2094 from a high-emission future run (ssp585), in which the radiative forcing increases by 8.5 W m$^{-2}$ by year 2100. Differences between the future and historical measures projected changes in a warmer climate. The specific variables used in this work include 6-hourly full-column air temperature ($T$), specific humidity ($q$), horizontal and meridional winds ($U$ and $V$), air pressures ($P$), as well as 6-hourly near-surface (2-m) air temperature ($T_s$) and specific humidity ($q_s$). Surface pressure ($P_s$) is additionally used to derive full-column geopotential height ($Z$) based on the hydrostatic equation. To characterize maximum near-surface moist heat, we first extract all variables at the time of the daily maximum moist heat (MSE$_s$), from which the annual maximum MSE$_s$ is identified at each land grid point for each year. The timing of the annual maximum MSE$_s$ defines the critical time (denoted by subscript ``c''), when moist heat peaks and deep convection potentially initiates to terminate it. For example, CAPE$_c$ and CIN$_c$ are critical CAPE and CIN at the time of annual maximum MSE$_s$. These annual maxima are then averaged over the 30-year period for each climate scenario to generate climatologies for each model. For the ensemble mean results presented, all fields are linearly interpolated to a common 1$^o\times$1$^o$ grid for each model before averaging across multiple models. 

To ensure the credibility of projections, we present in the main text results based on ensemble means of four relatively well-performing models, including CNRM-ESM2-1, CNRM-CM6-1, MPI-ESM1-2-LR, and MPI-ESM1-2-HR. These models are selected based on assessments of each model's performance against the ERA5 reanalysis data \cite{Hersbach_etal_2020} for the historical period during 1980--2009. Although all models exhibit biases in annual maximum MSE$_s$ and the associated convective instability (CAPE$_c$) and low-level energy inversion (MSE$^*_{max}$) (Supplementary Fig. \ref{fig_s01}), the midlatitude mean biases are relatively small in the selected models but substantially larger, especially for CAPE$_c$, in the remaining five models (Supplementary Fig. \ref{fig_s02}$A$). In addition, the selected models better reproduce spatial patterns of MSE$_s$ and CAPE$_c$ compared to the other models, as evidenced by their higher spatial correlations with patterns in the ERA5 reanalysis (Supplementary Fig. \ref{fig_s02}$B$). More importantly, the selected models capture well the inversion constraints on maximum moist heat (Supplementary Fig. \ref{fig_s02}$C$) and potential convection (Supplementary Fig. \ref{fig_s02}$D$), which provides a robust physical foundation for this work. These four models have also been identified in previous studies as among the best-performing CMIP6 models in terms of quantitatively reproducing convective environments and basic mean states over North America \cite{chavas_li_2022_biases}. 

While not the focus of the present study, it is worth noting the diverse departures from the inversion constraint relations across models (Supplementary Fig. \ref{fig_s02}$C$--$D$). Given the strong inter-model correlations between biases in MSE$_s$ and biases in CAPE$_c$ (Supplementary Fig. \ref{fig_s02}$A$) \cite{chavas_li_2022_biases}, and their intrinsic connections to low-level inversions \cite{Li_Tamarin-Brodsky_2025}, improved understanding of the processes driving biases in low-level inversions and in their constraining strength on heat and convection may offer valuable pathways for reducing modeling uncertainties of extreme heat and convection, which deserves further investigation in the future. We acknowledge recent bias-correction efforts for near-surface heat metrics in CMIP6 \cite{vecellio_etal_2023,kong_huber_2025_data}, but we stick with raw model outputs to preserve physical consistency among near-surface heat metrics, full-column convective parameters, and large-scale dynamical flow patterns. 

\subsection*{Composite analysis}

Extreme moist heat events are detected for sub-regions over central-northeastern North America and northeastern Asia, corresponding to emerging future hotspots of moist heat and potential convection in the midlatitudes. An event is identified as an extreme case if near-surface moist static energy (MSE$_s$) at all land grids within the region exceeds their respective annual 95th percentile values. The 95th percentile is defined for each grid point and each year based on the time series of daily maximum MSE$_s$. We apply the criterion on a daily basis, and when a day is identified as an extreme event, all associated variables are stored for composite analysis. In the main text, we focus on the U.S. Midwest (white boxes to the east in Fig. \ref{fig_02}$A$) and the western portion of northeastern Asia (i.e., eastern Mongolia and northern China; white boxes to the east in Fig. \ref{fig_02}$D$). Across all years and selected models, this eventually yields 330 versus 191 cases for the North America sub-region and 323 versus 347 cases for the northeastern Asia sub-region in the historical and future climate simulations, respectively. Ensemble means of all cases from each simulation and their differences (future minus historical) are then analyzed to build up the dynamical linkages shown in Fig. \ref{fig_03}$A$, $B$, $D$, and $E$. 

The composite soundings shown in Fig. \ref{fig_03}$C$ and $F$ represent mean static energy profiles evaluated at the time of annual maximum MSE$_s$ for land grids within each sub-region. Based on \cite{Li_Tamarin-Brodsky_2025}, the vertical profile of parcel buoyancy is alternatively determined by comparing the parcel's initial moist static energy (i.e., MSE$_s$) with the environmental static energy profile: 

\begin{equation}
\label{eq5}
b(z)\sim 
\begin{cases} 
MSE_{s}-(DSE(z)+L_vq_s), &z< LCL\\\\

MSE_{s}-MSE^{*}(z), &z\geq LCL
\end{cases}
\end{equation}
The environmental static energy profile is therefore defined as $DSE(z)+L_vq_s$ if $z<$ LCL and as $MSE^{*}(z)$ if $z\geq$ LCL, where DSE($z$) is the environmental dry static energy at a given height $z$ and LCL is the lifting condensation level for a surface-based air parcel. 

In the supplementary, we additionally show composite results for sub-regions over the lower Northeast of North America (Supplementary Fig. \ref{fig_s09}$A$--$B$), central Northeast of North America (Supplementary Fig. \ref{fig_s09}$C$--$D$), lower Northeast Asia (Supplementary Fig. \ref{fig_s09}$E$--$F$), and central Northeast Asia (Supplementary Fig. \ref{fig_s09}$G$--$H$). The composite patterns of extreme moist weather events are consistent across these sub-regions within central-northeastern North America hotspot (Fig. \ref{fig_03}$A$--$C$ and Supplementary Fig. \ref{fig_s09}$A$--$D$) and the northeastern Asia hotspot (Fig. \ref{fig_03}$D$--$F$ and Supplementary Fig. \ref{fig_s09}$E$--$H$). 

\subsection*{CESM experiments}

To examine the net impacts of orographic elevated heating in the midlatitudes, we conduct a control historical simulation ($CTL$) and compare it with an experiment in which elevated heating is removed ($EXP$), using the Community Atmosphere Model version 6 (CAM6) coupled to the Community Land Model version 5 (CLM5) within the Community Earth System Model version 2.1.5 (CESM2.1.5) \cite{Danabasoglu_etal_2020}.

The $CTL$ simulation uses the default FHIST component configuration within CESM2.1.5, configured on a 0.9$^o$ $\times$ 1.25$^o$ latitude-longitude grid mesh with 32 hybrid sigma-pressure levels. The global ocean is prescribed with monthly historical sea surface temperatures and ice coverage created from merged Hadley-NOAA/OI products \cite{Hurrell_etal_2008}. We integrate the simulation over the period 1979--2014 on NCAR’s Derecho supercomputers, discard the first 6 years for spinup, and analyze the 6-hourly output from 1985 to 2014. Previous studies have shown that comparable historical simulations using CESM2.1.0 \cite{Li_etal_2020} and CESM2.1.1 \cite{li_etal_2024} successfully reproduce realistic climatological patterns of severe convective environments, associated synoptic features including low-level jets and elevated inversions, and large-scale dynamic processes such as storm tracks and mid- to upper-level flows, despite a modest high bias in severe convective instability and near-surface moist static energy. Reader is referred to \cite{Li_etal_2020,li_etal_2024} for a comprehensive model assessment and validation. 

The $EXP$ simulation is integrated identically to $CTL$, except that surface albedo is increased to eliminate the heating source over midlatitude high terrains in the Northern Hemisphere. Specifically, both the direct and diffuse albedo of soil and vegetated surfaces are set to 0.98 (setting values to 1 causes model crushes) over high terrains where surface altitude exceeds 800 m between 15--70$^o$ N (excluding Greenland). Snow cover adjustments to albedo are retained, and as a result, albedo values in $EXP$ are not exactly 0.98 over highlands. Nevertheless, $EXP$ produces highland albedo values that are substantially higher than those in $CTL$ (Supplementary Fig. \ref{fig_s10}$A$--$C$). This modification effectively reduces solar heating over highlands, thereby eliminating orographic elevated heating in the midlatitudes (Supplementary Fig. \ref{fig_s10}$D$--$F$) and enabling explicit examination of downstream responses of concurrent moist heat and convection to the upstream elevated heating in the midlatitudes. This experimental design follows established modeling approaches used to isolate topographic thermal influences on monsoon precipitation in past studies \cite{boos_kuang_2010,boos_pascale_2021}. 

\newpage

\newpage

\newpage

\newpage


%


\clearpage 

%
\bibliography{all-bibtex} 

\begin{thebibliography}{100}
\providecommand{\url}[1]{\texttt{#1}}
\expandafter\ifx\csname urlstyle\endcsname\relax
  \providecommand{\doi}[1]{doi:\discretionary{}{}{}#1}\else
  \providecommand{\doi}{doi:\discretionary{}{}{}\begingroup \urlstyle{rm}\Url}\fi

\bibitem{morss_etal_2011}
R.~E. Morss, O.~V. Wilhelmi, G.~A. Meehl, L.~Dilling, Improving societal outcomes of extreme weather in a changing climate: an integrated perspective. \emph{Annual Review of Environment and Resources} \textbf{36}~(1), 1--25 (2011).

\bibitem{ummenhofer_etal_2017}
C.~C. Ummenhofer, G.~A. Meehl, Extreme weather and climate events with ecological relevance: a review. \emph{Philosophical Transactions of the Royal Society B: Biological Sciences} \textbf{372}~(1723), 20160135 (2017).

\bibitem{sherwood_huber_2010}
S.~C. Sherwood, M.~Huber, An adaptability limit to climate change due to heat stress. \emph{Proceedings of the National Academy of Sciences} \textbf{107}~(21), 9552--9555 (2010).

\bibitem{baldwin_etal_2023}
J.~W. Baldwin, \emph{et~al.}, Humidity’s role in heat-related health outcomes: a heated debate. \emph{Environmental health perspectives} \textbf{131}~(5), 055001 (2023).

\bibitem{stevens_2005atmospheric}
B.~Stevens, Atmospheric moist convection. \emph{Annu. Rev. Earth Planet. Sci.} \textbf{33}~(1), 605--643 (2005).

\bibitem{bao_etal_2024}
J.~Bao, B.~Stevens, L.~Kluft, C.~Muller, Intensification of daily tropical precipitation extremes from more organized convection. \emph{Science Advances} \textbf{10}~(8), eadj6801 (2024).

\bibitem{coumou_Rahmstorf_2012}
D.~Coumou, S.~Rahmstorf, A decade of weather extremes. \emph{Nature climate change} \textbf{2}~(7), 491--496 (2012).

\bibitem{matthews_2018}
T.~Matthews, Humid heat and climate change. \emph{Progress in Physical Geography: Earth and Environment} \textbf{42}~(3), 391--405 (2018).

\bibitem{zipser_liu_2021}
E.~J. Zipser, C.~Liu, Extreme convection vs. extreme rainfall: A global view. \emph{Current Climate Change Reports} \textbf{7}~(4), 121--130 (2021).

\bibitem{kang_eltahir_2018}
S.~Kang, E.~A. Eltahir, North China Plain threatened by deadly heatwaves due to climate change and irrigation. \emph{Nature communications} \textbf{9}~(1), 2894 (2018).

\bibitem{zhang_prein_2025}
Z.~Zhang, \emph{et~al.}, Moisture from US Corn Belt fuels more intense convective storms. \emph{Communications Earth \& Environment}  (2025).

\bibitem{kong_huber_2023}
Q.~Kong, M.~Huber, Regimes of Soil Moisture--Wet-Bulb Temperature Coupling with Relevance to Moist Heat Stress. \emph{Journal of Climate} \textbf{36}~(22), 7925--7942 (2023).

\bibitem{li_etal_2024}
F.~Li, D.~R. Chavas, B.~Medeiros, K.~A. Reed, K.~L. Rasmussen, Upstream surface roughness and terrain are strong drivers of contrast in tornado potential between North and South America. \emph{Proceedings of the National Academy of Sciences} \textbf{121}~(26), e2315425121 (2024).

\bibitem{tamarin_etal_2020}
T.~Tamarin-Brodsky, K.~Hodges, B.~J. Hoskins, T.~G. Shepherd, Changes in Northern Hemisphere temperature variability shaped by regional warming patterns. \emph{Nature Geoscience} \textbf{13}~(6), 414--421 (2020).

\bibitem{jiang_etal_2025}
Q.~Jiang, D.~T. Dawson~II, F.~Li, D.~R. Chavas, Classifying synoptic patterns driving tornadic storms and associated spatial trends in the United States. \emph{npj Climate and Atmospheric Science} \textbf{8}~(1), 7 (2025).

\bibitem{raymond_etal_2020}
C.~Raymond, T.~Matthews, R.~M. Horton, The emergence of heat and humidity too severe for human tolerance. \emph{Science Advances} \textbf{6}~(19), eaaw1838 (2020).

\bibitem{rogers_etal_2021}
C.~D. Rogers, \emph{et~al.}, Recent increases in exposure to extreme humid-heat events disproportionately affect populated regions. \emph{Geophysical Research Letters} \textbf{48}~(19), e2021GL094183 (2021).

\bibitem{domeisen_etal_2023}
D.~I. Domeisen, \emph{et~al.}, Prediction and projection of heatwaves. \emph{Nature Reviews Earth \& Environment} \textbf{4}~(1), 36--50 (2023).

\bibitem{barriopedro_etal_2023}
D.~Barriopedro, R.~Garc{\'\i}a-Herrera, C.~Ord{\'o}nez, D.~G. Miralles, S.~Salcedo-Sanz, Heat waves: Physical understanding and scientific challenges. \emph{Reviews of Geophysics} \textbf{61}~(2), e2022RG000780 (2023).

\bibitem{romps_lu_2022}
D.~M. Romps, Y.-C. Lu, Chronically underestimated: a reassessment of US heat waves using the extended heat index. \emph{Environmental Research Letters} \textbf{17}~(9), 094017 (2022).

\bibitem{lanzante_2024new}
J.~R. Lanzante, A new heat stress index for climate change assessment. \emph{Bulletin of the American Meteorological Society} \textbf{105}~(12), E2482--E2495 (2024).

\bibitem{xu_etal_2025}
W.~R. Xu, K.~W. Dixon, N.~Zenes, D.~Adams-Smith, Sometimes missing the heat: the risk of underestimating extreme heat days with daily maximum heat index approximation. \emph{International Journal of Biometeorology} pp. 1--15 (2025).

\bibitem{Zipser_etal_2006}
E.~J. Zipser, D.~J. Cecil, C.~Liu, S.~W. Nesbitt, D.~P. Yorty, Where are the most intense thunderstorms on Earth? \emph{Bulletin of the American Meteorological Society} \textbf{87}~(8), 1057--1072 (2006).

\bibitem{Brooks_etal_2003}
H.~E. Brooks, J.~W. Lee, J.~P. Craven, The spatial distribution of severe thunderstorm and tornado environments from global reanalysis data. \emph{Atmospheric Research} \textbf{67}, 79--94 (2003), \doi{https://doi.org/10.1016/S0169-8095(03)00045-0}.

\bibitem{prein_2023}
A.~F. Prein, Thunderstorm straight line winds intensify with climate change. \emph{Nature Climate Change} \textbf{13}~(12), 1353--1359 (2023).

\bibitem{tang_etal_2019}
B.~H. Tang, V.~A. Gensini, C.~R. Homeyer, Trends in United States large hail environments and observations. \emph{NPJ Climate and Atmospheric Science} \textbf{2}~(1), 45 (2019).

\bibitem{BillionDollar_2025}
N.~N.~C. for Environmental Information~(NCEI), U.S. Billion-Dollar Weather and Climate Disasters (2025), \doi{10.25921/stkw-7w73}, \url{https://www.ncei.noaa.gov/access/billions/}.

\bibitem{sauter_etal_2023a}
C.~Sauter, \emph{et~al.}, Compound extreme hourly rainfall preconditioned by heatwaves most likely in the mid-latitudes. \emph{Weather and Climate Extremes} \textbf{40}, 100563 (2023).

\bibitem{treppiedi_etal_2024}
D.~Treppiedi, G.~Villarini, L.~V. Noto, Climate change exacerbates the compounding of heat stress and flooding in the mid-latitudes. \emph{International Journal of Climatology} \textbf{44}~(7), 2283--2296 (2024).

\bibitem{sauter_2023c}
C.~Sauter, C.~J. White, H.~J. Fowler, S.~Westra, Temporally compounding heatwave--heavy rainfall events in Australia. \emph{International Journal of Climatology} \textbf{43}~(2), 1050--1061 (2023).

\bibitem{xu_etal_2023mega}
Z.~Xu, Y.~Zhang, G.~Bl{\"o}schl, S.~Piao, Mega forest fires intensify flood magnitudes in southeast Australia. \emph{Geophysical Research Letters} \textbf{50}~(12), e2023GL103812 (2023).

\bibitem{ganguli_etal_2025}
P.~Ganguli, B.~Merz, Increasing probability of extreme rainfall preconditioned by humid heatwaves in global coastal megacities. \emph{npj Climate and Atmospheric Science} \textbf{8}~(1), 144 (2025).

\bibitem{sauter_etal_2023b}
C.~Sauter, J.~L. Catto, H.~J. Fowler, S.~Westra, C.~J. White, Compounding Heatwave-Extreme Rainfall Events Driven by Fronts, High Moisture, and Atmospheric Instability. \emph{Journal of Geophysical Research: Atmospheres} \textbf{128}~(21), e2023JD038761 (2023).

\bibitem{vyshkvarkova_etal_2022}
E.~Vyshkvarkova, O.~Sukhonos, Compound extremes of air temperature and precipitation in Eastern Europe. \emph{Climate} \textbf{10}~(9), 133 (2022).

\bibitem{you_wang_2021}
J.~You, S.~Wang, Higher probability of occurrence of hotter and shorter heat waves followed by heavy rainfall. \emph{Geophysical Research Letters} \textbf{48}~(17), e2021GL094831 (2021).

\bibitem{li_etal_2022heat}
C.~Li, \emph{et~al.}, Substantial increase in heavy precipitation events preceded by moist heatwaves over China during 1961--2019. \emph{Frontiers in Environmental Science} \textbf{10}, 951392 (2022).

\bibitem{miao_etal_2024}
L.~Miao, \emph{et~al.}, Unveiling the dynamics of sequential extreme precipitation-heatwave compounds in China. \emph{npj Climate and Atmospheric Science} \textbf{7}~(1), 67 (2024).

\bibitem{zhang_Villarini_2020}
W.~Zhang, G.~Villarini, Deadly compound heat stress-flooding hazard across the Central United States. \emph{Geophysical Research Letters} \textbf{47}~(15), e2020GL089185 (2020).

\bibitem{hao_etal_2024}
Y.~Hao, \emph{et~al.}, Evaluating the effects of heatwave events on hydrological processes in the contiguous United States (2003--2022). \emph{Journal of Hydrology} \textbf{637}, 131368 (2024).

\bibitem{zhou_etal_2024heat}
Z.~Zhou, \emph{et~al.}, Global increase in future compound heat stress-heavy precipitation hazards and associated socio-ecosystem risks. \emph{Npj Climate and Atmospheric Science} \textbf{7}~(1), 33 (2024).

\bibitem{liu_etal_2025}
J.~Liu, J.~Chen, J.~Yin, A.~D. King, Time of emergence of record-shattering compound heatwave-extreme precipitation events and their socio-economic exposures. \emph{Geophysical Research Letters} \textbf{52}~(16), e2025GL116884 (2025).

\bibitem{russo_etal_2017}
S.~Russo, J.~Sillmann, A.~Sterl, Humid heat waves at different warming levels. \emph{Scientific reports} \textbf{7}~(1), 7477 (2017).

\bibitem{Lepore_etal_2021}
C.~Lepore, R.~Abernathey, N.~Henderson, J.~T. Allen, M.~K. Tippett, {Future Global Convective Environments in CMIP6 Models}. \emph{Earth's Future} \textbf{9}~(12), e2021EF002277 (2021).

\bibitem{jin_etal_2025}
H.~Jin, \emph{et~al.}, Global escalation of more frequent and intense compound heatwave-extreme precipitation events. \emph{EGUsphere} \textbf{2025}, 1--30 (2025).

\bibitem{coffel_etal_2017}
E.~D. Coffel, R.~M. Horton, A.~De~Sherbinin, Temperature and humidity based projections of a rapid rise in global heat stress exposure during the 21st century. \emph{Environmental Research Letters} \textbf{13}~(1), 014001 (2017).

\bibitem{vecellio_etal_2023}
D.~J. Vecellio, Q.~Kong, W.~L. Kenney, M.~Huber, Greatly enhanced risk to humans as a consequence of empirically determined lower moist heat stress tolerance. \emph{Proceedings of the National Academy of Sciences} \textbf{120}~(42), e2305427120 (2023).

\bibitem{raymond_etal_2025}
C.~Raymond, L.~Suarez-Gutierrez, V.~Thompson, K.~van~der Wiel, Distinct favored regions for historical record-setting and future record-breaking humid heat. \emph{AGU Advances} \textbf{6}~(6), e2025AV001963 (2025).

\bibitem{Gensini_Brooks_2018}
V.~A. Gensini, H.~E. Brooks, {Spatial trends in United States tornado frequency}. \emph{npj Climate and Atmospheric Science} \textbf{1}~(1), 38 (2018).

\bibitem{Battaglioli_etal_2025}
F.~Battaglioli, M.~Taszarek, P.~Groenemeijer, T.~Púčik, A.~R. Rädler, Contrasting trends in very large hail events and related economic losses across the globe. \emph{Nature Geoscience}  (2025), \doi{https://doi.org/10.1038/s41561-025-01868-0}.

\bibitem{zhang_etal_2021_heat}
Y.~Zhang, I.~Held, S.~Fueglistaler, Projections of tropical heat stress constrained by atmospheric dynamics. \emph{Nature Geoscience} \textbf{14}~(3), 133--137 (2021).

\bibitem{Singh_etal_2017}
M.~S. Singh, Z.~Kuang, E.~D. Maloney, W.~M. Hannah, B.~O. Wolding, Increasing potential for intense tropical and subtropical thunderstorms under global warming. \emph{Proceedings of the National Academy of Sciences} \textbf{114}~(44), 11657--11662 (2017), \doi{https://doi.org/10.1073/pnas.1707603114}.

\bibitem{da_etal_2025}
N.~A. Da~Silva, J.~O. Haerter, Super-Clausius--Clapeyron scaling of extreme precipitation explained by shift from stratiform to convective rain type. \emph{Nature Geoscience}  (2025).

\bibitem{wang_moyer_2023}
Z.~Wang, E.~J. Moyer, Robust relationship between midlatitudes CAPE and moist static energy surplus in present and future simulations. \emph{Geophysical Research Letters} \textbf{50}~(14), e2023GL104163 (2023).

\bibitem{zhang_2002convective}
G.~J. Zhang, Convective quasi-equilibrium in midlatitude continental environment and its effect on convective parameterization. \emph{Journal of Geophysical Research: Atmospheres} \textbf{107}~(D14), ACL--12 (2002).

\bibitem{lafleur_etal_2023}
A.~T. LaFleur, R.~L. Tanamachi, D.~T. Dawson, D.~D. Turner, Factors affecting the rapid recovery of CAPE on 31 March 2016 during VORTEX-Southeast. \emph{Monthly Weather Review} \textbf{151}~(6), 1459--1477 (2023).

\bibitem{sillmann_etal_2017}
J.~Sillmann, \emph{et~al.}, Understanding, modeling and predicting weather and climate extremes: Challenges and opportunities. \emph{Weather and climate extremes} \textbf{18}, 65--74 (2017).

\bibitem{shaw_bjorn_2025}
T.~A. Shaw, B.~Stevens, The other climate crisis. \emph{Nature} \textbf{639}~(8056), 877--887 (2025).

\bibitem{Li_Tamarin-Brodsky_2025}
F.~Li, T.~Tamarin-Brodsky, Atmospheric stability sets maximum moist heat and convection in the midlatitudes. \emph{Science Advances} \textbf{12}~(1), eaea8453 (2026).

\bibitem{buzan_huber_2020}
J.~R. Buzan, M.~Huber, Moist heat stress on a hotter Earth. \emph{Annual Review of Earth and Planetary Sciences} \textbf{48}~(1), 623--655 (2020).

\bibitem{zhang_boos_2023}
Y.~Zhang, W.~R. Boos, An upper bound for extreme temperatures over midlatitude land. \emph{Proceedings of the National Academy of Sciences} \textbf{120}~(12), e2215278120 (2023).

\bibitem{duan_etal_2024}
S.~Q. Duan, F.~Ahmed, J.~D. Neelin, Moist heatwaves intensified by entrainment of dry air that limits deep convection. \emph{Nature Geoscience} \textbf{17}~(9), 837--844 (2024).

\bibitem{emanuel_etal_1994}
K.~A. Emanuel, J.~David~Neelin, C.~S. Bretherton, On large-scale circulations in convecting atmospheres. \emph{Quarterly Journal of the Royal Meteorological Society} \textbf{120}~(519), 1111--1143 (1994).

\bibitem{neelin_zeng_2000}
J.~D. Neelin, N.~Zeng, A quasi-equilibrium tropical circulation model—Formulation. \emph{Journal of the atmospheric sciences} \textbf{57}~(11), 1741--1766 (2000).

\bibitem{singh_ogorman_2013}
M.~S. Singh, P.~A. O'Gorman, Influence of entrainment on the thermal stratification in simulations of radiative-convective equilibrium. \emph{Geophysical Research Letters} \textbf{40}~(16), 4398--4403 (2013).

\bibitem{Doswell_2001}
C.~A. Doswell, Severe convective storms—An overview, in \emph{Severe convective storms} (Springer), pp. 1--26 (2001).

\bibitem{Tuckman_etal_2023}
P.~Tuckman, V.~Agard, K.~Emanuel, Evolution of convective energy and inhibition before instances of large CAPE. \emph{Monthly Weather Review} \textbf{151}~(1), 321--338 (2023).

\bibitem{Tuckman_Emanuel_2024}
P.~Tuckman, K.~Emanuel, Origins of Extreme CAPE around the World. \emph{Journal of Geophysical Research: Atmospheres} \textbf{129}~(22), e2024JD041833 (2024).

\bibitem{wood_Bretherton_2006}
R.~Wood, C.~S. Bretherton, On the relationship between stratiform low cloud cover and lower-tropospheric stability. \emph{Journal of climate} \textbf{19}~(24), 6425--6432 (2006).

\bibitem{klein_etal_2018}
S.~A. Klein, A.~Hall, J.~R. Norris, R.~Pincus, Low-cloud feedbacks from cloud-controlling factors: A review. \emph{Shallow clouds, water vapor, circulation, and climate sensitivity} pp. 135--157 (2018).

\bibitem{Eyring_etal_2016}
V.~Eyring, \emph{et~al.}, {Overview of the Coupled Model Intercomparison Project Phase 6 (CMIP6) experimental design and organization}. \emph{Geoscientific Model Development} \textbf{9}~(5), 1937--1958 (2016), \doi{10.5194/gmd-9-1937-2016}, \url{https://gmd.copernicus.org/articles/9/1937/2016/}.

\bibitem{Carlson_etal_1983}
T.~N. Carlson, S.~G. Benjamin, G.~S. Forbes, Y.~F. Li, Elevated mixed layers in the regional severe storm environment: Conceptual model and case studies. \emph{Monthly Weather Review} \textbf{111}~(7), 1453--1474 (1983), \doi{https://doi.org/10.1175/1520-0493(1983)111<1453:EMLITR>2.0.CO;2}.

\bibitem{Li_etal_2021}
F.~Li, D.~R. Chavas, K.~A. Reed, N.~Rosenbloom, D.~T. Dawson~II, The role of elevated terrain and the Gulf of Mexico in the production of severe local storm environments over North America. \emph{Journal of Climate} \textbf{34}~(19), 7799--7819 (2021).

\bibitem{raymond_etal_2021}
C.~Raymond, \emph{et~al.}, On the controlling factors for globally extreme humid heat. \emph{Geophysical Research Letters} \textbf{48}~(23), e2021GL096082 (2021).

\bibitem{Chen_etal_2020}
J.~Chen, A.~Dai, Y.~Zhang, K.~L. Rasmussen, Changes in convective available potential energy and convective inhibition under global warming. \emph{Journal of Climate} \textbf{33}~(6), 2025--2050 (2020).

\bibitem{Taszarek_etal_2021_global}
M.~Taszarek, J.~T. Allen, M.~Marchio, H.~E. Brooks, Global climatology and trends in convective environments from ERA5 and rawinsonde data. \emph{NPJ climate and atmospheric science} \textbf{4}, 1--11 (2021).

\bibitem{ashley_etal_2023}
W.~S. Ashley, A.~M. Haberlie, V.~A. Gensini, The future of supercells in the United States. \emph{Bulletin of the American Meteorological Society} \textbf{104}~(1), E1--E21 (2023).

\bibitem{hill_etal_2025}
S.~A. Hill, \emph{et~al.}, More extreme Indian monsoon rainfall in El Ni{\~n}o summers. \emph{Science} \textbf{389}~(6766), 1220--1224 (2025).

\bibitem{schumacher_Rasmussen_2020}
R.~S. Schumacher, K.~L. Rasmussen, The formation, character and changing nature of mesoscale convective systems. \emph{Nature Reviews Earth \& Environment} \textbf{1}~(6), 300--314 (2020).

\bibitem{bishop_etal_2021}
D.~A. Bishop, \emph{et~al.}, Placing the east-west North American aridity gradient in a multi-century context. \emph{Environmental Research Letters} \textbf{16}~(11), 114043 (2021).

\bibitem{you_etal_2022}
Q.~You, \emph{et~al.}, Recent frontiers of climate changes in East Asia at global warming of 1.5° C and 2° C. \emph{Npj Climate and Atmospheric Science} \textbf{5}~(1), 80 (2022).

\bibitem{palazzi_etal_2017}
E.~Palazzi, L.~Filippi, J.~von Hardenberg, Insights into elevation-dependent warming in the Tibetan Plateau-Himalayas from CMIP5 model simulations. \emph{Climate Dynamics} \textbf{48}~(11), 3991--4008 (2017).

\bibitem{pepin_etal_2025}
N.~Pepin, \emph{et~al.}, Elevation-dependent climate change in mountain environments. \emph{Nature Reviews Earth \& Environment} pp. 1--17 (2025).

\bibitem{rangwala_etal_2016}
I.~Rangwala, E.~Sinsky, J.~R. Miller, Variability in projected elevation dependent warming in boreal midlatitude winter in CMIP5 climate models and its potential drivers. \emph{Climate Dynamics} \textbf{46}~(7), 2115--2122 (2016).

\bibitem{minder_etal_2018}
J.~R. Minder, T.~W. Letcher, C.~Liu, The character and causes of elevation-dependent warming in high-resolution simulations of Rocky Mountain climate change. \emph{Journal of Climate} \textbf{31}~(6), 2093--2113 (2018).

\bibitem{lian_etal_2021}
X.~Lian, \emph{et~al.}, Multifaceted characteristics of dryland aridity changes in a warming world. \emph{Nature Reviews Earth \& Environment} \textbf{2}~(4), 232--250 (2021).

\bibitem{cook_etal_2020}
B.~I. Cook, \emph{et~al.}, Twenty-first century drought projections in the CMIP6 forcing scenarios. \emph{Earth's Future} \textbf{8}~(6), e2019EF001461 (2020).

\bibitem{mckinnon_etal_2024}
K.~A. McKinnon, I.~R. Simpson, A.~P. Williams, The pace of change of summertime temperature extremes. \emph{Proceedings of the National Academy of Sciences} \textbf{121}~(42), e2406143121 (2024).

\bibitem{bauer_etal_2025}
A.~M. Bauer, L.~R. Vargas~Zeppetello, C.~Proistosescu, An Analytical Model for the Influence of Soil Moisture on Temperature Extremes in the Midlatitudes. \emph{Journal of Climate} \textbf{38}~(24), 7395--7413 (2025).

\bibitem{zhang_etal_2025moisture}
Z.~Zhang, \emph{et~al.}, Moisture from US Corn Belt fuels more intense convective storms. \emph{Communications Earth \& Environment}  (2025).

\bibitem{Li_Chavas_2021}
F.~Li, D.~R. Chavas, Midlatitude Continental CAPE Is Predictable From Large-Scale Environmental Parameters. \emph{Geophysical Research Letters} \textbf{48}~(8), e2020GL091799 (2021).

\bibitem{Emanuel_2023}
K.~Emanuel, On the physics of high CAPE. \emph{Journal of the Atmospheric Sciences} \textbf{80}~(11), 2669--2683 (2023).

\bibitem{li_yanai_1996}
C.~Li, M.~Yanai, The onset and interannual variability of the Asian summer monsoon in relation to land--sea thermal contrast. \emph{Journal of Climate} \textbf{9}~(2), 358--375 (1996).

\bibitem{boos_kuang_2010}
W.~R. Boos, Z.~Kuang, Dominant control of the South Asian monsoon by orographic insulation versus plateau heating. \emph{Nature} \textbf{463}~(7278), 218--222 (2010).

\bibitem{boos_pascale_2021}
W.~R. Boos, S.~Pascale, Mechanical forcing of the North American monsoon by orography. \emph{Nature} \textbf{599}~(7886), 611--615 (2021).

\bibitem{lu_etal_2018}
M.~Lu, \emph{et~al.}, Possible effect of the Tibetan Plateau on the “upstream” climate over west Asia, North Africa, south Europe and the North Atlantic. \emph{Climate Dynamics} \textbf{51}~(4), 1485--1498 (2018).

\bibitem{chen_etal_2021TP}
Z.~Chen, Q.~Wen, H.~Yang, Impact of Tibetan Plateau on North African precipitation. \emph{Climate Dynamics} \textbf{57}~(9), 2767--2777 (2021).

\bibitem{rasmussen_etal_2020}
K.~L. Rasmussen, A.~F. Prein, R.~M. Rasmussen, K.~Ikeda, C.~Liu, Changes in the convective population and thermodynamic environments in convection-permitting regional climate simulations over the United States. \emph{Climate Dynamics} \textbf{55}~(1), 383--408 (2020).

\bibitem{chernokulsky_etal_2022}
A.~Chernokulsky, \emph{et~al.}, Atmospheric severe convective events in Russia: Changes observed from different data. \emph{Russian Meteorology and Hydrology} \textbf{47}~(5), 343--354 (2022).

\bibitem{cui_etal_2024}
W.~Cui, T.~J. Galarneau~Jr, K.~A. Hoogewind, Changes in mesoscale convective system precipitation structures in response to a warming climate. \emph{Journal of Geophysical Research: Atmospheres} \textbf{129}~(9), e2023JD039920 (2024).

\bibitem{dai_etal_2024}
P.~Dai, J.~Nie, Y.~Yu, R.~Wu, Constraints on regional projections of mean and extreme precipitation under warming. \emph{Proceedings of the National Academy of Sciences} \textbf{121}~(11), e2312400121 (2024).

\bibitem{Agee_etal_2016}
E.~Agee, J.~Larson, S.~Childs, A.~Marmo, Spatial redistribution of {U.S.} Tornado activity between 1954 and 2013. \emph{Journal of Applied Meteorology and Climatology} \textbf{55}~(8), 1681--1697 (2016), \doi{https://doi.org/10.1175/JAMC-D-15-0342.1}.

\bibitem{feldmann_etal_2025}
M.~Feldmann, \emph{et~al.}, European supercell thunderstorms—A prevalent current threat and an increasing future hazard. \emph{Science Advances} \textbf{11}~(35), eadx0513 (2025).

\bibitem{rothlisberger_Papritz_2023}
M.~R{\"o}thlisberger, L.~Papritz, Quantifying the physical processes leading to atmospheric hot extremes at a global scale. \emph{Nature Geoscience} \textbf{16}~(3), 210--216 (2023).

\bibitem{weckwerth_Parsons_2006}
T.~M. Weckwerth, D.~B. Parsons, A review of convection initiation and motivation for IHOP\_2002. \emph{Monthly weather review} \textbf{134}~(1), 5--22 (2006).

\bibitem{bennett_etal_2006review}
L.~J. Bennett, K.~A. Browning, A.~M. Blyth, D.~J. Parker, P.~A. Clark, A review of the initiation of precipitating convection in the United Kingdom. \emph{Quarterly Journal of the Royal Meteorological Society: A journal of the atmospheric sciences, applied meteorology and physical oceanography} \textbf{132}~(617), 1001--1020 (2006).

\bibitem{clement_etal_2009}
A.~C. Clement, R.~Burgman, J.~R. Norris, Observational and model evidence for positive low-level cloud feedback. \emph{Science} \textbf{325}~(5939), 460--464 (2009).

\bibitem{zhou_etal_2016}
C.~Zhou, M.~D. Zelinka, S.~A. Klein, Impact of decadal cloud variations on the Earth’s energy budget. \emph{Nature Geoscience} \textbf{9}~(12), 871--874 (2016).

\bibitem{medeiros_etal_2011}
B.~Medeiros, C.~Deser, R.~A. Tomas, J.~E. Kay, Arctic inversion strength in climate models. \emph{Journal of Climate} \textbf{24}~(17), 4733--4740 (2011).

\bibitem{li_mace_2023}
X.~Li, G.~G. Mace, C.~Strong, S.~K. Krueger, Wintertime Cooling of the Arctic TOA by Low-Level Clouds. \emph{Geophysical Research Letters} \textbf{50}~(17), e2023GL104869 (2023).

\bibitem{largeron_staquet_2016}
Y.~Largeron, C.~Staquet, Persistent inversion dynamics and wintertime PM10 air pollution in Alpine valleys. \emph{Atmospheric Environment} \textbf{135}, 92--108 (2016).

\bibitem{sun_etal_2025inversion}
M.~Sun, Z.~Xie, X.~Yao, S.~Wang, L.~Dong, Multilayer temperature inversion structures and their potential impact on atmospheric pollution in northwest China. \emph{Atmospheric Environment} \textbf{343}, 120998 (2025).

\bibitem{Ribeiro_Bosart_2018}
B.~Z. Ribeiro, L.~F. Bosart, Elevated mixed layers and associated severe thunderstorm environments in {South and North Americas}. \emph{Monthly Weather Review} \textbf{146}~(1), 3--28 (2018), \doi{https://doi.org/10.1175/MWR-D-17-0121.1}.

\bibitem{baldwin_vecchi_2016}
J.~Baldwin, G.~Vecchi, Influence of the Tian Shan on arid extratropical Asia. \emph{Journal of Climate} \textbf{29}~(16), 5741--5762 (2016).

\bibitem{xie_etal_2025}
Y.~Xie, \emph{et~al.}, Changing Northern Hemisphere weather linked to warming amplification in High Mountain Asia. \emph{Communications Earth \& Environment} \textbf{6}~(1), 932 (2025).

\bibitem{neal_etal_2022}
E.~Neal, C.~S. Huang, N.~Nakamura, The 2021 Pacific Northwest heat wave and associated blocking: Meteorology and the role of an upstream cyclone as a diabatic source of wave activity. \emph{Geophysical Research Letters} \textbf{49}~(8), e2021GL097699 (2022).

\bibitem{madden_Robitaille_1970}
R.~Madden, F.~Robitaille, A comparison of the equivalent potential temperature and the static energy. \emph{Journal of the Atmospheric Sciences} \textbf{27}~(2), 327--329 (1970).

\bibitem{betts_1974}
A.~K. Betts, Further comments on “A comparison of the equivalent potential temperature and the static energy”. \emph{Journal of the Atmospheric Sciences} \textbf{31}~(6), 1713--1715 (1974).

\bibitem{seferian_etal_2019}
R.~S{\'e}f{\'e}rian, \emph{et~al.}, Evaluation of CNRM Earth system model, CNRM-ESM2-1: Role of Earth system processes in present-day and future climate. \emph{Journal of Advances in Modeling Earth Systems} \textbf{11}~(12), 4182--4227 (2019).

\bibitem{voldoire_etal_2019}
A.~Voldoire, \emph{et~al.}, Evaluation of CMIP6 deck experiments with CNRM-CM6-1. \emph{Journal of Advances in Modeling Earth Systems} \textbf{11}~(7), 2177--2213 (2019).

\bibitem{mauritsen_etal_2019}
T.~Mauritsen, \emph{et~al.}, Developments in the MPI-M Earth System Model version 1.2 (MPI-ESM1. 2) and its response to increasing CO2. \emph{Journal of Advances in Modeling Earth Systems} \textbf{11}~(4), 998--1038 (2019).

\bibitem{muller_etal_2018}
W.~A. M{\"u}ller, \emph{et~al.}, A higher-resolution version of the max planck institute earth system model (MPI-ESM1. 2-HR). \emph{Journal of Advances in Modeling Earth Systems} \textbf{10}~(7), 1383--1413 (2018).

\bibitem{cherchi_etal_2019}
A.~Cherchi, \emph{et~al.}, Global mean climate and main patterns of variability in the CMCC-CM2 coupled model. \emph{Journal of Advances in Modeling Earth Systems} \textbf{11}~(1), 185--209 (2019).

\bibitem{yukimoto_etal_2019}
S.~Yukimoto, \emph{et~al.}, The Meteorological Research Institute Earth System Model version 2.0, MRI-ESM2. 0: Description and basic evaluation of the physical component. \emph{Journal of the Meteorological Society of Japan. Ser. II} \textbf{97}~(5), 931--965 (2019).

\bibitem{tatebe_etal_2019}
H.~Tatebe, \emph{et~al.}, Description and basic evaluation of simulated mean state, internal variability, and climate sensitivity in MIROC6. \emph{Geoscientific Model Development} \textbf{12}~(7), 2727--2765 (2019).

\bibitem{hajima_etal_2020}
T.~Hajima, \emph{et~al.}, Development of the MIROC-ES2L Earth system model and the evaluation of biogeochemical processes and feedbacks. \emph{Geoscientific Model Development} \textbf{13}~(5), 2197--2244 (2020).

\bibitem{swart_etal_2019}
N.~C. Swart, \emph{et~al.}, The Canadian earth system model version 5 (CanESM5. 0.3). \emph{Geoscientific Model Development} \textbf{12}~(11), 4823--4873 (2019).

\bibitem{Hersbach_etal_2020}
H.~Hersbach, \emph{et~al.}, The {ERA5} global reanalysis. \emph{Quarterly Journal of the Royal Meteorological Society} \textbf{146}~(730), 1999--2049 (2020).

\bibitem{chavas_li_2022_biases}
D.~R. Chavas, F.~Li, Biases in CMIP6 historical US severe convective storm environments driven by biases in mean-state near-surface moist static energy. \emph{Geophysical Research Letters} \textbf{49}~(23), e2022GL098527 (2022).

\bibitem{kong_huber_2025_data}
Q.~Kong, M.~Huber, A global high-resolution and bias-corrected dataset of CMIP6 projected heat stress metrics. \emph{Scientific Data} \textbf{12}~(1), 246 (2025).

\bibitem{Danabasoglu_etal_2020}
G.~Danabasoglu, \emph{et~al.}, {The community earth system model version 2 (CESM2)}. \emph{Journal of Advances in Modeling Earth Systems} \textbf{12}~(2), e2019MS001916 (2020).

\bibitem{Hurrell_etal_2008}
J.~W. Hurrell, J.~J. Hack, D.~Shea, J.~M. Caron, J.~Rosinski, {A new sea surface temperature and sea ice boundary dataset for the Community Atmosphere Model}. \emph{Journal of Climate} \textbf{21}~(19), 5145--5153 (2008).

\bibitem{Li_etal_2020}
F.~Li, D.~R. Chavas, K.~A. Reed, D.~T. Dawson~II, Climatology of severe local storm environments and synoptic-scale features over North America in ERA5 reanalysis and CAM6 simulation. \emph{Journal of Climate} \textbf{33}~(19), 8339--8365 (2020).

\bibitem{NCAR_RDA_ERA5}
{European Centre for Medium-Range Weather Forecasts}, {ERA5 Reanalysis (0.25 Degree Latitude-Longitude Grid)} (2019), \url{https://doi.org/10.5065/BH6N-5N20}.

\bibitem{May_etal_2022}
R.~M. May, \emph{et~al.}, {MetPy: A meteorological Python library for data analysis and visualization}. \emph{Bulletin of the American Meteorological Society} \textbf{103}~(10), E2273--E2284 (2022).

\end{thebibliography}
\bibliographystyle{sciencemag}

%


\clearpage 

\section*{Acknowledgments}
We thank Paul O'Gorman and Kerry Emanuel for valuable discussions. We acknowledge the use of Derecho and Casper supercomputers provided by NSF National Center for Atmospheric Research.
\paragraph*{Funding:}
This research is part of the MIT Climate Grand Challenge on Weather and Climate Extremes. F.L. and T.T-B. acknowledge supports provided by Schmidt Sciences.
\paragraph*{Author contributions:}
Conceptualization: F.L. and T.T-B. Methodology: F.L. and T.T-B. Investigation: F.L. Formal analysis: F.L. Validation: F.L. Visualization: F.L. Software: F.L. Data curation: F.L. Writing-original draft: F.L. Writing-review and editing: F.L. and T.T-B. Funding acquisition: T.T-B. Supervision: T.T-B.
\paragraph*{Competing interests:}
All authors declare they have no competing interests.
\paragraph*{Data and materials availability:}
The 6-hourly model-level CMIP6 historical and ssp585 experimental data are publicly available at \url{https://esgf-node.llnl.gov/search/cmip6} \cite{Eyring_etal_2016}. The 6-hourly near-surface and pressure-level ERA5 data during 1980--2022 are publicly available at \url{https://rda.ucar.edu/datasets/ds633.0} \cite{NCAR_RDA_ERA5}. CESM2.1.5 model code is publicly available at \url{https://www.cesm.ucar.edu/models/cesm2/release_download.html} \cite{Danabasoglu_etal_2020}. The xcape and Metpy python packages for calculating convective variables are publicly available at \url{https://github.com/xgcm/xcape} \cite{Lepore_etal_2021} and \url{https://unidata.github.io/MetPy} \cite{May_etal_2022}, respectively.


\subsection*{Supplementary materials} 
Figs. S1 to S10\\ 


\newpage


\renewcommand{\thefigure}{S\arabic{figure}}
\renewcommand{\thetable}{S\arabic{table}}
\renewcommand{\theequation}{S\arabic{equation}}
\renewcommand{\thepage}{S\arabic{page}}
\setcounter{figure}{0}
\setcounter{table}{0}
\setcounter{equation}{0}
\setcounter{page}{1} 


\begin{center}
\section*{Supplementary Materials for\\ \scititle}

Funing~Li$^{\ast}$,
Talia~Tamarin-Brodsky\\ 
\small$^\ast$Corresponding author. Email: lifuning1991@gmail.com
\end{center}

\subsubsection*{This PDF file includes:} 
Figures S1 to S10\\


%


\clearpage
\begin{figure} 
\centering
\centerline{\includegraphics[width=1.0\linewidth]{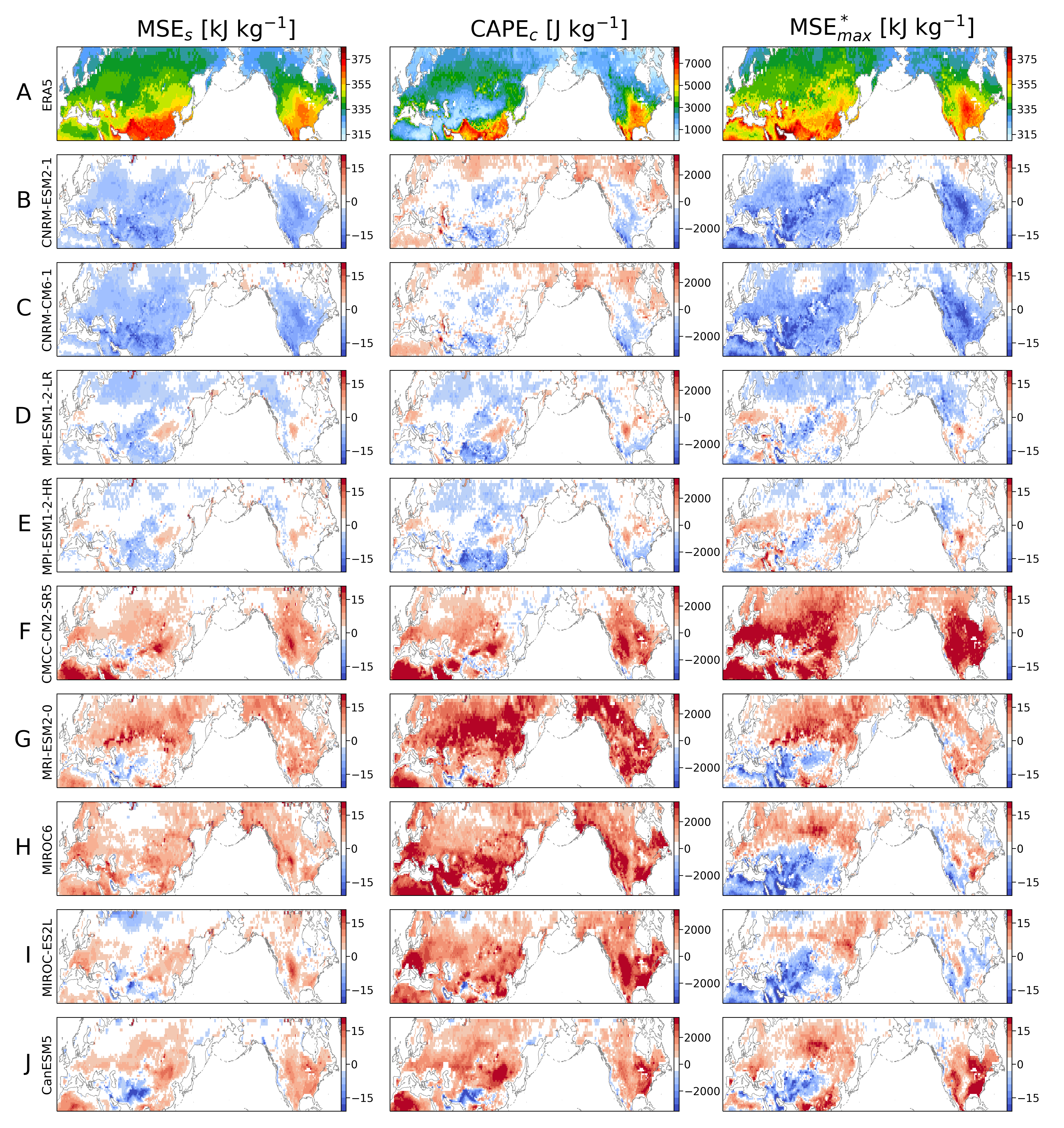}}  
\caption{\textbf{Assessment of model performance.} 
(\textbf{A}): Annual maximum MSE$_s$ (left), CAPE$_c$ (middle), and MSE$_{max}^{*}$ (right) from ERA5 reanalysis during 1980--2009. (\textbf{B}--\textbf{J}): Biases in maximum MSE$_s$, CAPE$_c$, and MSE$_{max}^{*}$ for each individual model relative to ERA5 (i.e., model minus ERA5) over 1980--2009.
}
\label{fig_s01} 
\end{figure}

\clearpage
\begin{figure} 
\centering
\centerline{\includegraphics[width=0.85\linewidth]{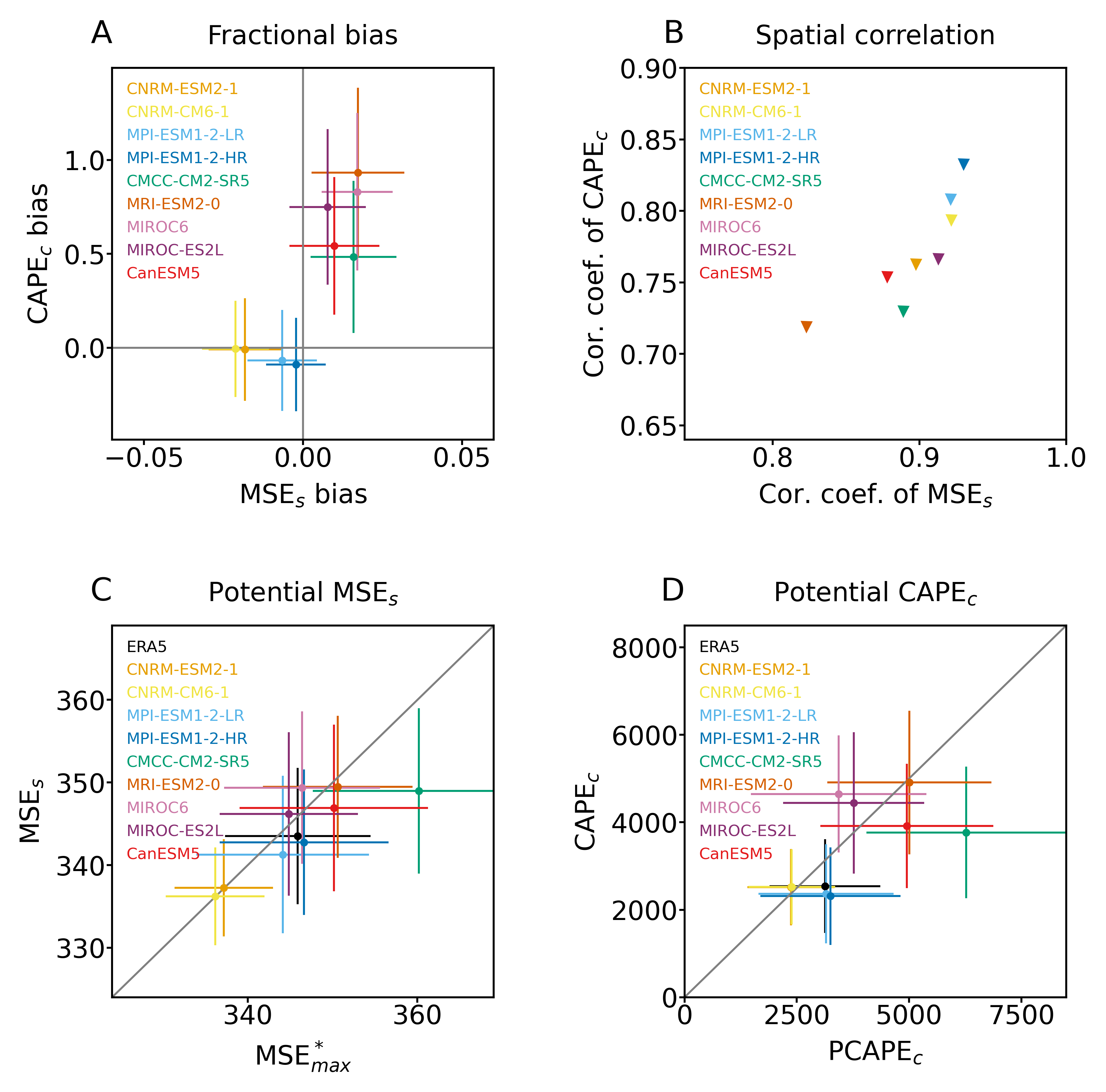}} 
\caption{\textbf{Assessment of model performance.} 
(\textbf{A}): Biases in annual maximum MSE$_s$ versus CAPE$_c$ averaged over midlatitude land during 1980--2009 for each model (model minus ERA5), expressed as fractions relative to ERA5 mean values, with bars indicating one standard deviation. (\textbf{B}): Spatial correlation coefficients over midlatitude land between each model and ERA5 for annual maximum MSE$_s$ versus CAPE$_c$. (\textbf{C}): Potential MSE$_s$ (i.e., MSE$_{max}^{*}$) versus MSE$_s$ at its annual maxima, averaged over midlatitude land during 1980--2009 for ERA5 and each model, with bars indicating one standard deviation. (\textbf{D}): Potential CAPE$_c$ (i.e., PCAPE$_c$) versus CAPE$_c$, averaged over midlatitude land during 1980--2009 for ERA5 and each model, with bars indicating one standard deviation.  
}
\label{fig_s02}  
\end{figure}

\clearpage
\begin{figure} 
\centering
\centerline{\includegraphics[width=1.0\linewidth]{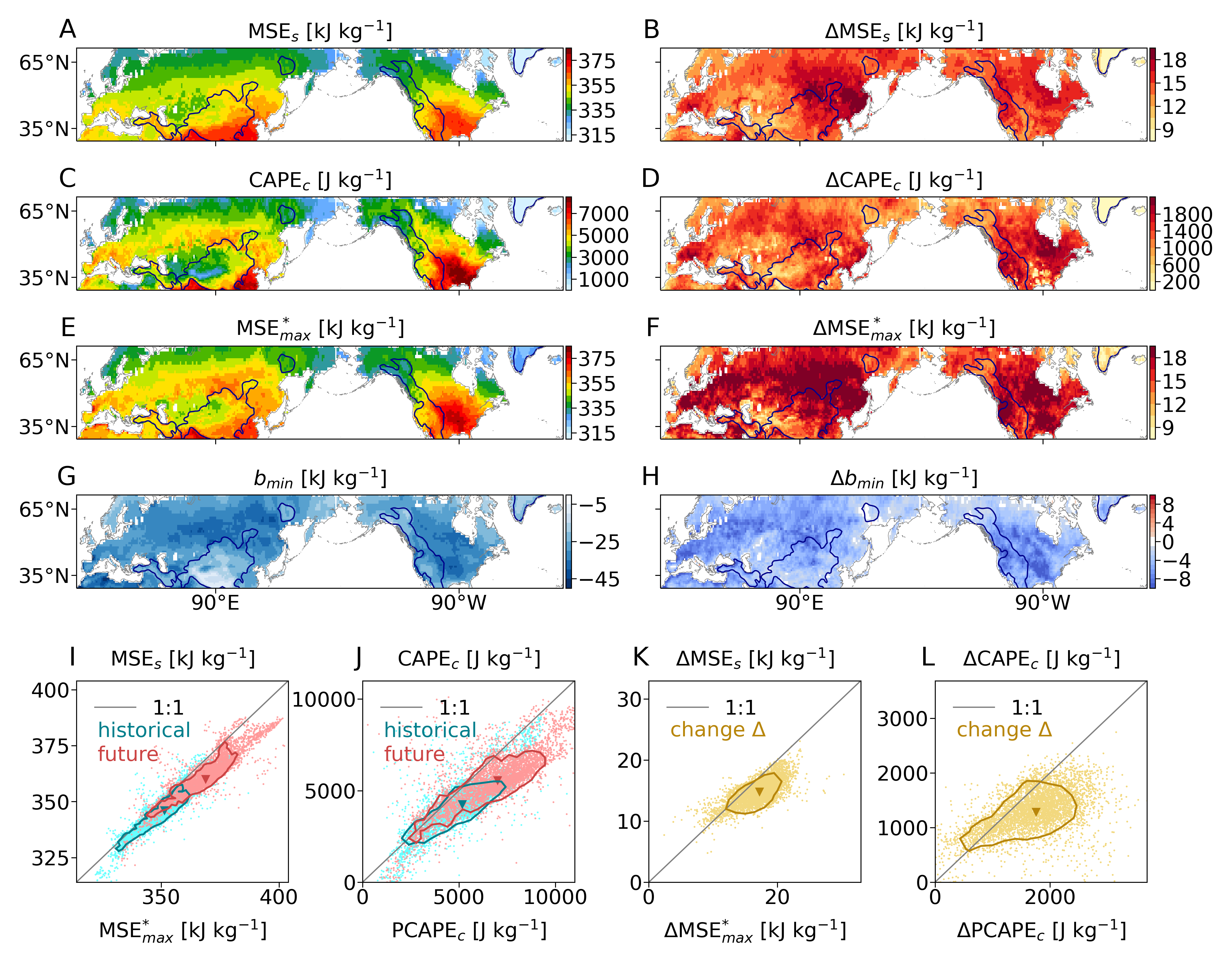}}
\caption{\textbf{Inversion constraints on concurrent moist heat and convection projections.} 
Same as Fig. \ref{fig_01} but for means of three relatively high-biased models: CMCC-CM2-SR5, MRI-ESM2-0, and MIROC6.  
}
\label{fig_s03}
\end{figure}

\clearpage
\begin{figure} 
\centering
\centerline{\includegraphics[width=1.0\linewidth]{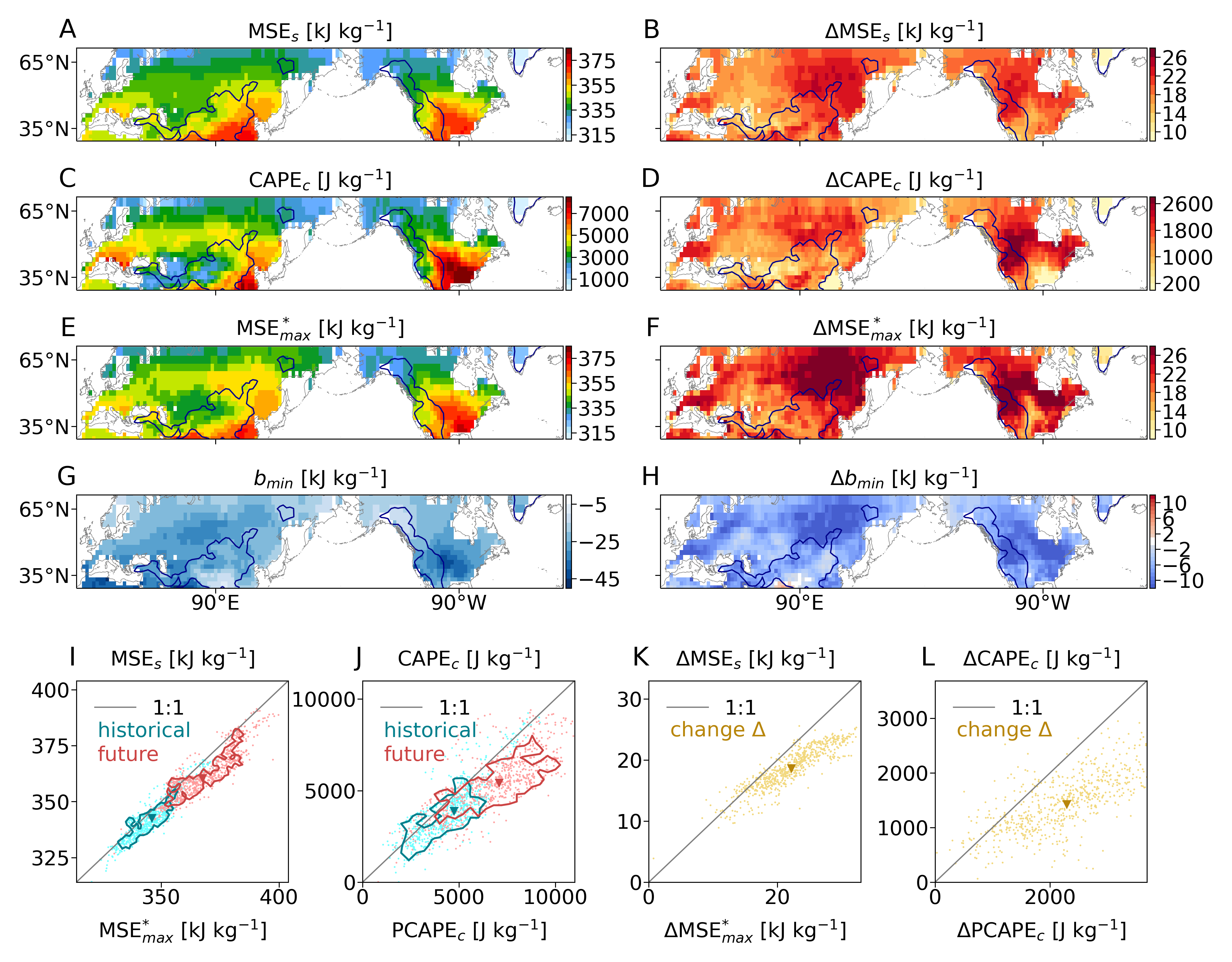}}
\caption{\textbf{Inversion constraints on concurrent moist heat and convection projections.} 
Same as Fig. \ref{fig_01} but for means of two low-resolution models: MIROC-ES2L and CanESM5. 
}
\label{fig_s04}
\end{figure}

\clearpage
\begin{figure} 
\centering
\centerline{\includegraphics[width=1.0\linewidth]{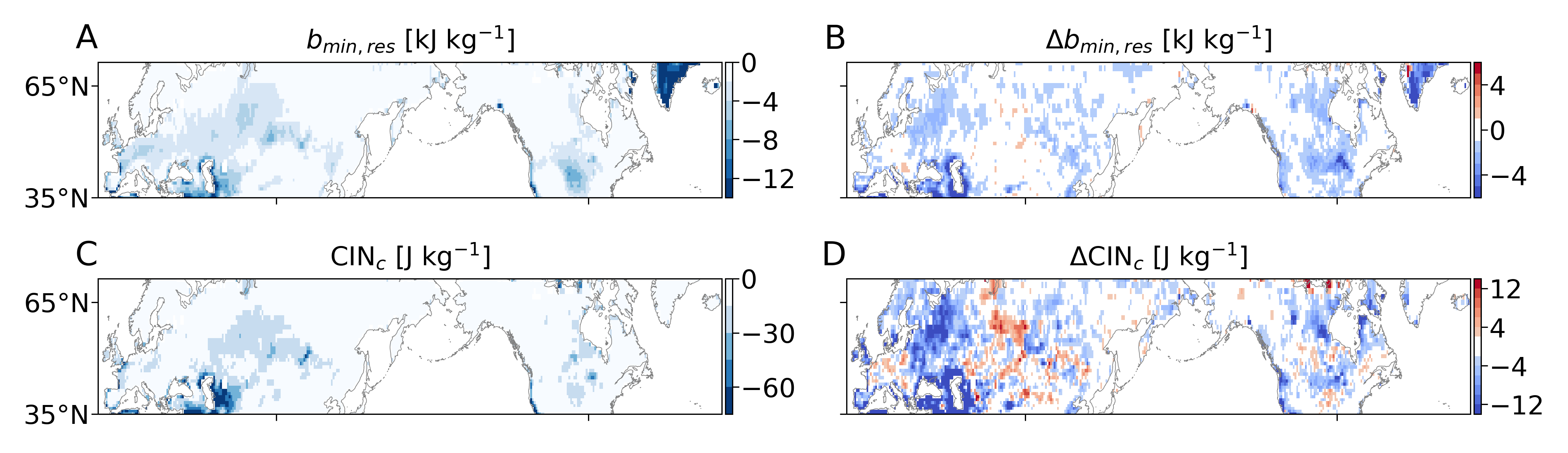}}
\caption{\textbf{Projected changes in residuals of inversion-induced energy barrier.} 
(\textbf{A}): Residual negative buoyancy of a near-surface air parcel ($b_{min,res}$), defined as the difference between MSE$_s$ and MSE$_{max}^*$ at the time of the annual maximum MSE$_s$, averaged over the historical period (1980--2009), and (\textbf{B}): their projected changes ($\Delta b_{min,res}$; 2065--2094 minus historical). (\textbf{C}--\textbf{D}): Same as (A--B) but for critical convective inhibition (CIN$_c$) at the time of annual maximum MSE$_s$.
} 
\label{fig_s05}
\end{figure} 

\clearpage
\begin{figure} 
\centering
\centerline{\includegraphics[width=1.0\linewidth]{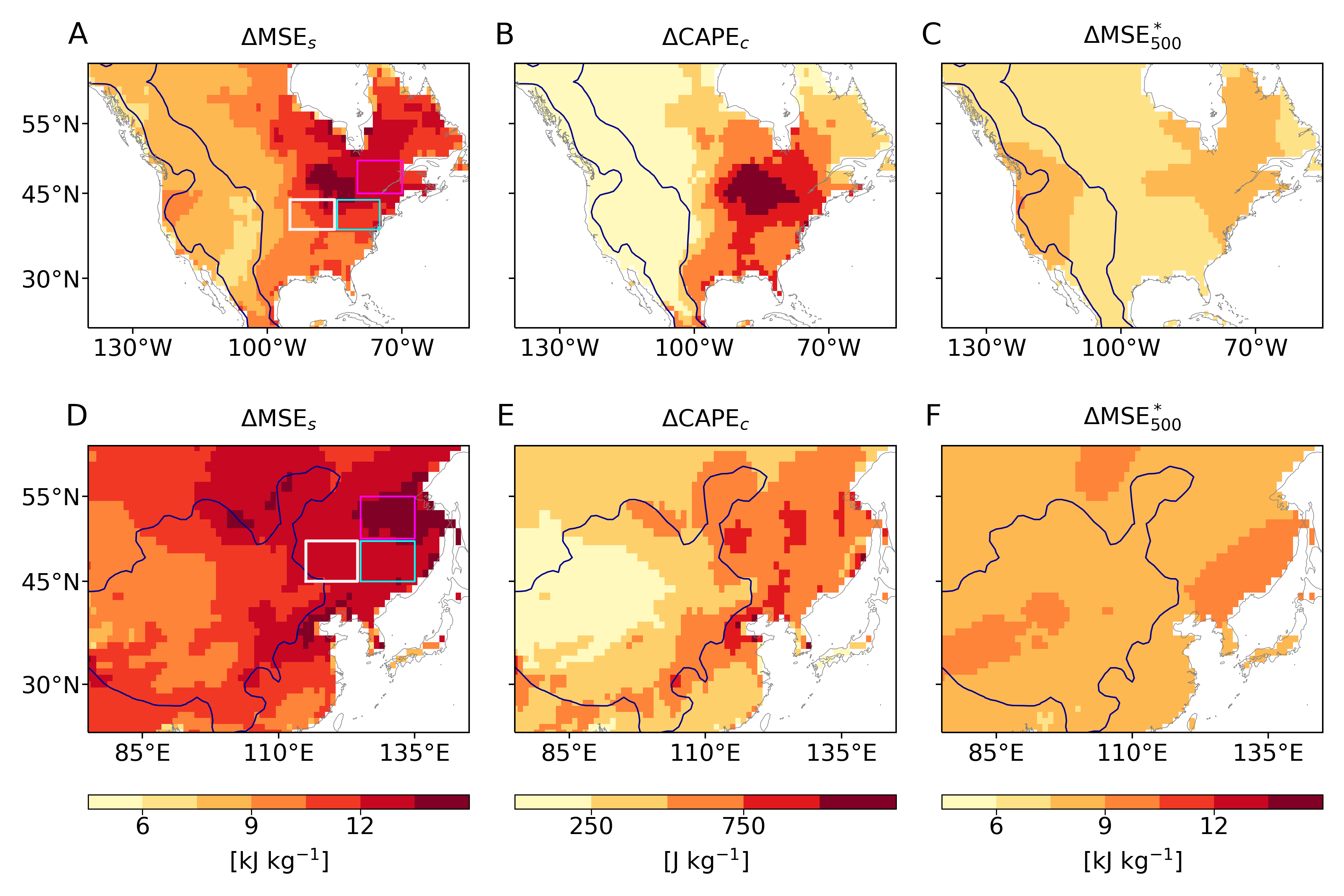}}
\caption{\textbf{Composite heat and convection extremes.} 
Projected changes ($\Delta$; future minus historical) in composite
(\textbf{A}) MSE$_s$,  (\textbf{B}) CAPE$_c$, and (\textbf{C}) 500-hPa saturated moist static energy (MSE$^*_{500}$) associated with extreme moist heat events over the U.S. Midwest (white box in (A)). (\textbf{D}--\textbf{F}): Same as (A--C) but associated with extreme moist heat events over eastern Mongolia and northern China (white box in (D)). Other boxes in (A) and (D) denote further sub-region analyses in Supplementary Fig. \ref{fig_s09}. 
}  
\label{fig_s06}
\end{figure}

\clearpage
\begin{figure} 
\centering
\centerline{\includegraphics[width=0.8\linewidth]{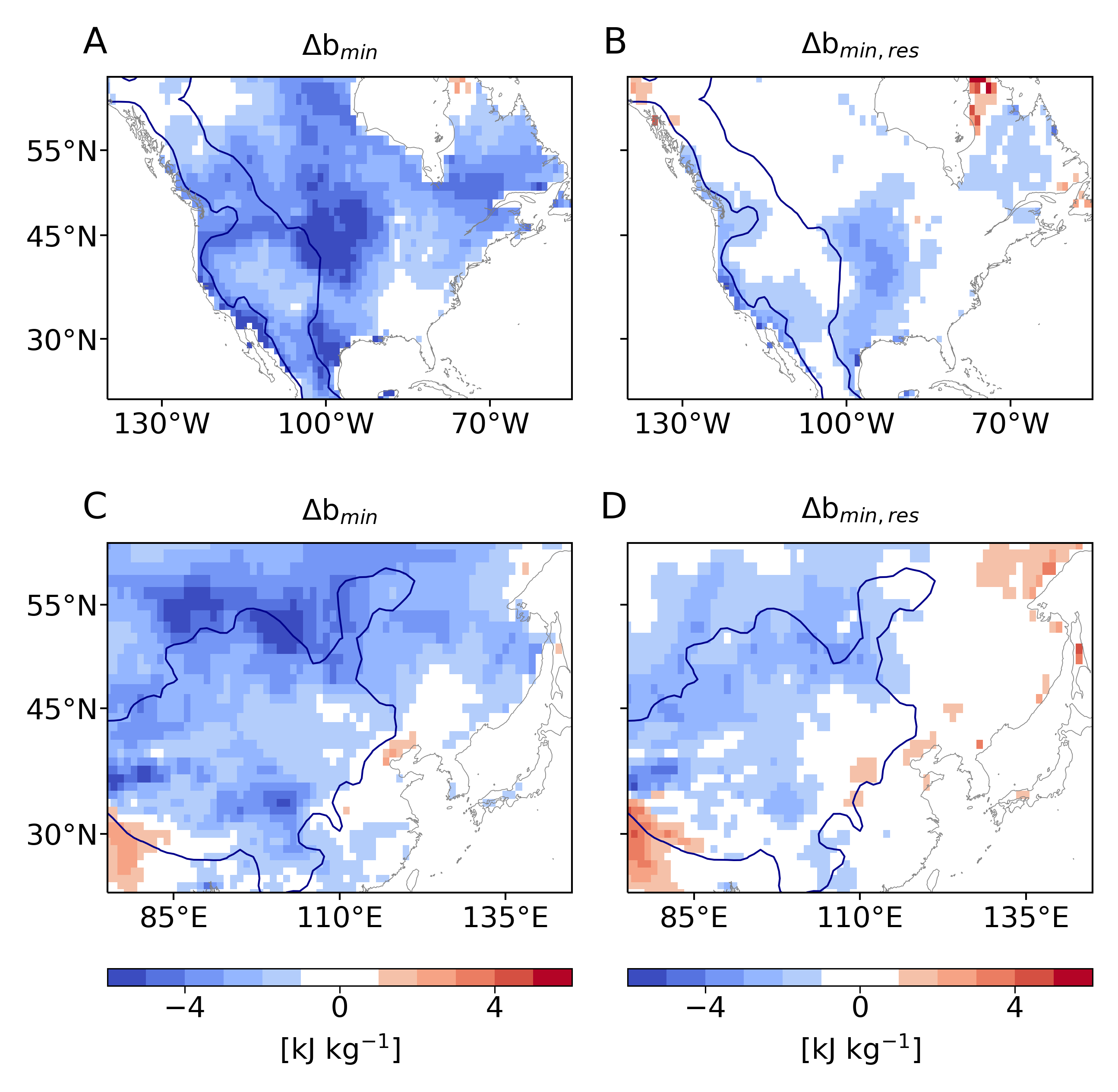}}
\caption{\textbf{Composite inversion-induced energy barriers.} 
(\textbf{A}--\textbf{B}): Projected changes ($\Delta$; future minus historical) associated with extreme moist heat events over the U.S. Midwest for (A) pre-existing inversion-induced energy barrier defined as the largest difference between MSE$_s$ and MSE$_{max}^*$ within five days prior annual maximum MSE$_s$, and (B) the residual inversion barrier defined as the difference between MSE$_s$ and MSE$_{max}^*$ at the time of the annual maximum MSE$_s$. (\textbf{C}--\textbf{D}): Same as (A--B) but associated with extreme moist heat events over eastern Mongolia and northern China. 
}  
\label{fig_s07}
\end{figure}

\clearpage
\begin{figure} 
\centering
\centerline{\includegraphics[width=0.8\linewidth]{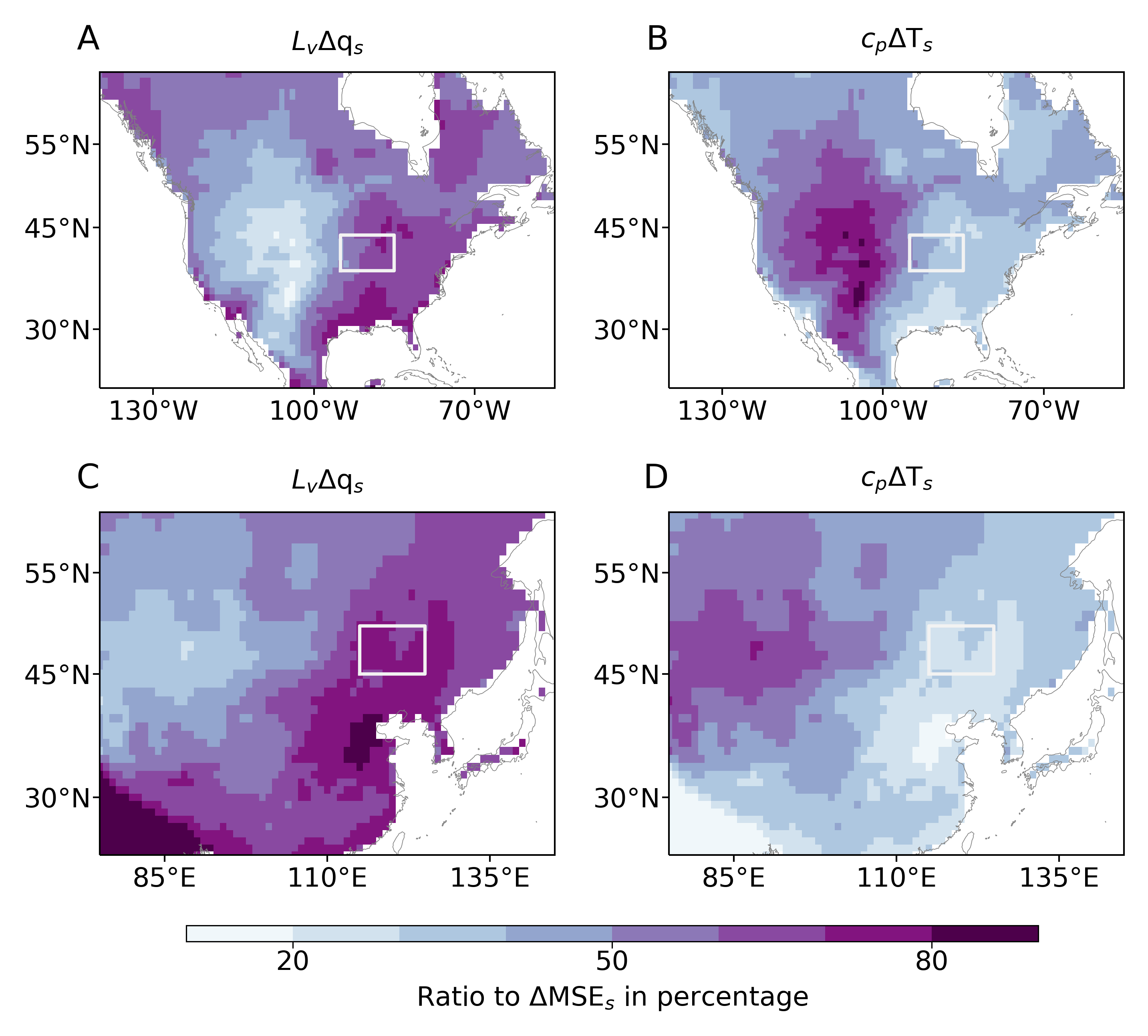}}
\caption{\textbf{Composite low-level temperature and humidity conditions}
(\textbf{A}--\textbf{B}): Projected changes ($\Delta$; future minus historical) associated with extreme moist heat events over the U.S. Midwest (white box) for (A) near-surface latent heat energy ($L_vq_s$; filled colors) and low-level winds at 925 hPa (vectors), and (B) near-surface sensible heat energy ($c_pT_s$). (\textbf{C}--\textbf{D}): Same as (A--B) but associated with extreme moist heat events over eastern Mongolia and northern China (white box). 
}  
\label{fig_s08}
\end{figure}

\clearpage
\begin{figure} 
\centering
\centerline{\includegraphics[width=0.8\linewidth]{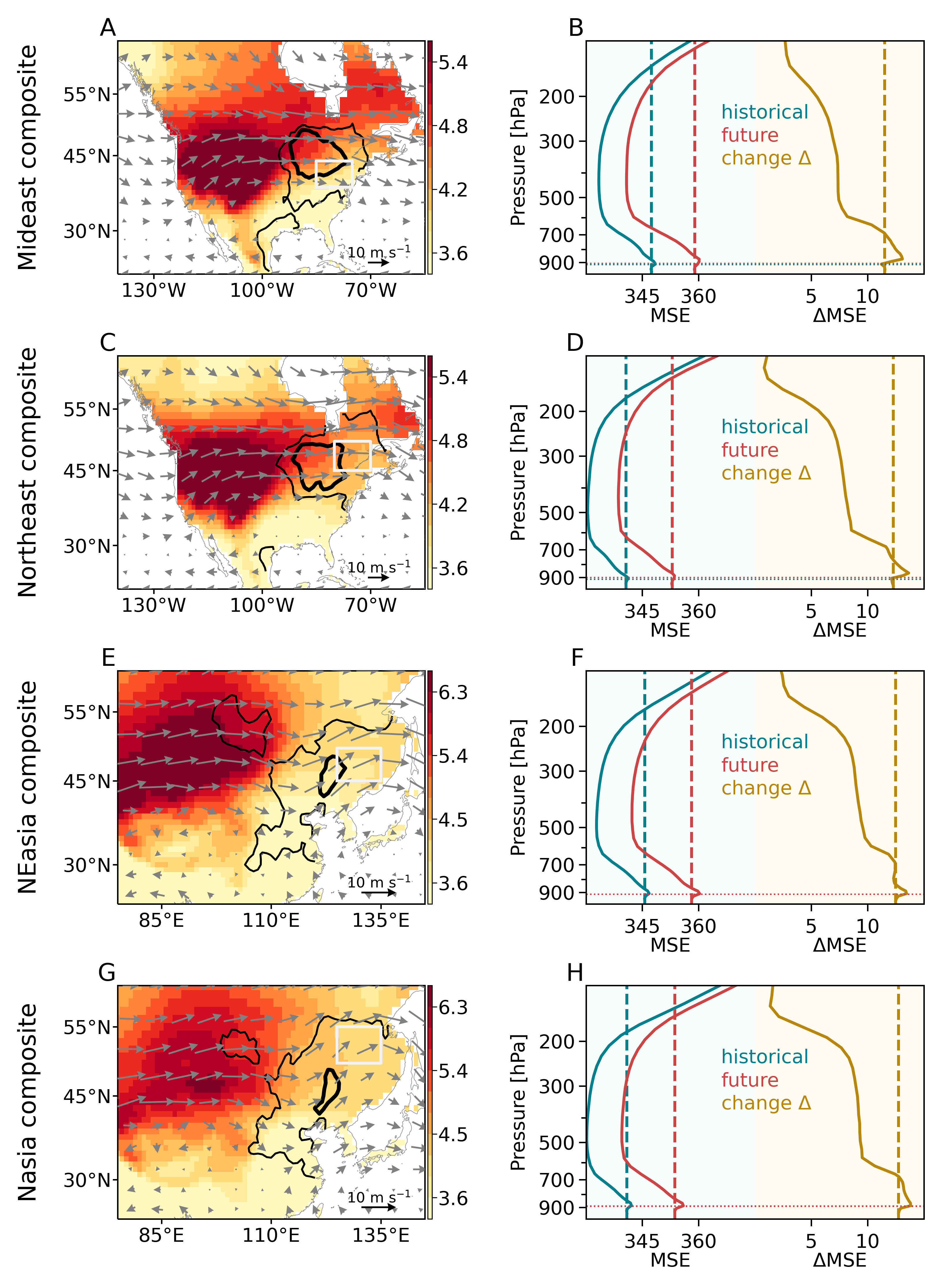}}
\caption{\textbf{Dynamical origins from upstream highland warming.} 
Same as Fig. \ref{fig_03} but associated with extreme moist heat events over (\textbf{A}--\textbf{B}) the lower Northeast of North America, (\textbf{C}--\textbf{D}) the central Northeast of North America, (\textbf{E}--\textbf{F}) the lower Northeast Asia, and (\textbf{G}--\textbf{H}) the central Northeast Asia. Sub-regions are indicated by white boxes.
} 
\label{fig_s09}
\end{figure}

\clearpage
\begin{figure} 
\centering
\centerline{\includegraphics[width=1.0\linewidth]{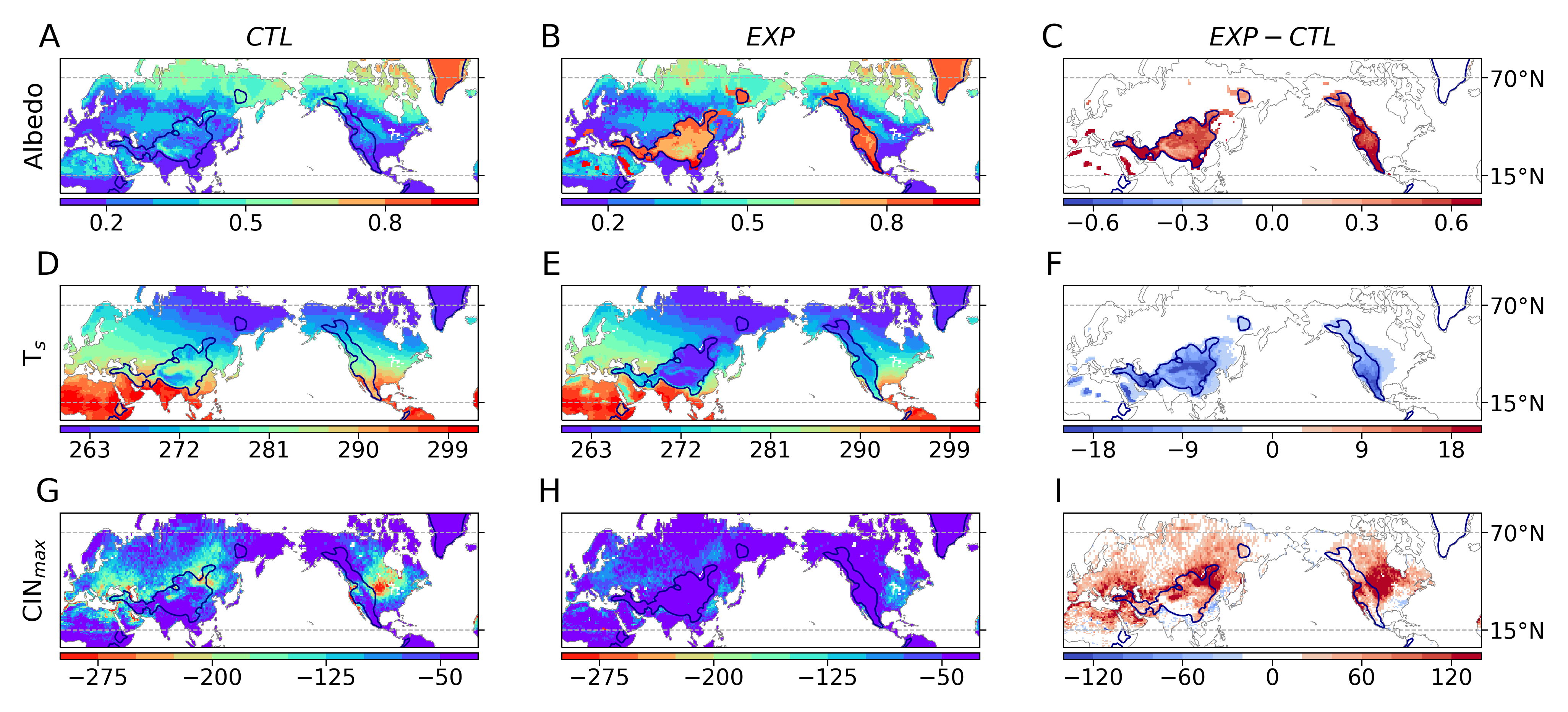}}
\caption{\textbf{Experimental design.} 
(\textbf{A}--\textbf{C}): Annual mean albedo from (A) the CESM2 historical control simulation ($CTL$), (B) the experiment with removed elevated heating ($EXP$), and (C) difference between $EXP$ and $CTL$. (\textbf{D}--\textbf{I}): Same as (A--C) but for (D--F) annual mean near-surface air temperature ($T_s$) and (G--I) the most negative convective inhibition (CIN$_{max}$) within five days prior annual maximum MSE$_s$.
}  
\label{fig_s10}
\end{figure}



\end{document}